\documentclass[10pt,twocolumn,letterpaper]{article}

\usepackage[pagenumbers]{wacv} % To force page numbers, e.g. for an arXiv version

\usepackage{graphicx}
\usepackage{amsmath}
\usepackage{amssymb}
\usepackage{booktabs}
\usepackage{amsfonts}       % blackboard math symbols
\usepackage{nicefrac}       % compact symbols for 1/2, etc.
\usepackage{microtype}      % microtypography       % colors
\usepackage{epsfig}
\usepackage{graphicx}
\usepackage{amsmath}
\usepackage{amssymb}
\usepackage{multirow}
\usepackage[font=small]{caption}
\usepackage{booktabs}
\usepackage{multirow}
\usepackage{setspace}
\usepackage{algorithm}
\usepackage[noend]{algpseudocode}
\usepackage{mathtools}
\usepackage{enumitem}
\usepackage{mathptmx}
\usepackage[table,xcdraw]{xcolor}

\usepackage[accsupp]{axessibility} % Improves PDF readability for those with disabilities
\definecolor{mygray}{HTML}{EFEFEF}

\usepackage[pagebackref,breaklinks,colorlinks]{hyperref}

\usepackage[capitalize]{cleveref}
\crefname{section}{Sec.}{Secs.}
\Crefname{section}{Section}{Sections}
\Crefname{table}{Table}{Tables}
\crefname{table}{Tab.}{Tabs.}

\def\wacvPaperID{2057} % *** Enter the WACV Paper ID here
\def\confName{WACV}
\def\confYear{2025}

\begin{document}

\title{MVAD: A Multiple Visual Artifact Detector for Video Streaming}

\author{%
  Chen Feng$\dagger$, Duolikun Danier$\dagger$, Fan Zhang$\dagger$, Alex Mackin$\ddagger$, Andrew Collins$\ddagger$, David Bull$\dagger$ \\
  $\dagger$ Visual Information Lab, University of Bristol, BS1 5DD, United Kingdom \\
  $\ddagger$Amazon Prime Video, 1 Principal Place, Worship Street, London, EC2A 2FA, United Kingdom\\
  \texttt{\{chen.feng, duolikun.danier, fan.zhang, dave.bull\}@bristol.ac.uk}, \\ \texttt{\{acmackin, accllin\}@amazon.co.uk}
}
\maketitle

%%%%%%%%% ABSTRACT
\begin{abstract}
   Visual artifacts are often introduced into streamed video content, due to prevailing conditions during content production and delivery. Since these can degrade the quality of the user's experience, it is important to automatically and accurately detect them in order to enable effective quality measurement and enhancement. Existing detection methods often focus on a single type of artifact and/or determine the presence of an artifact through thresholding objective quality indices. Such approaches have been reported to offer inconsistent prediction performance and are also impractical for real-world applications where multiple artifacts co-exist and interact. In this paper, we propose a Multiple Visual Artifact Detector, MVAD, for video streaming which, for the first time, is able to detect multiple artifacts using a single framework that is not reliant on video quality assessment models. Our approach employs a new Artifact-aware Dynamic Feature Extractor (ADFE) to obtain artifact-relevant spatial features within each frame for multiple artifact types. The extracted features are further processed by a Recurrent Memory Vision Transformer (RMViT) module, which captures both short-term and long-term temporal information within the input video. The proposed network architecture is optimized in an end-to-end manner based on a new, large and diverse training database that is generated by simulating the video streaming pipeline and based on Adversarial Data Augmentation. This model has been evaluated on two video artifact databases, Maxwell and BVI-Artifact, and achieves consistent and improved prediction results for ten target visual artifacts when compared to seven existing single and multiple artifact detectors. The source code and training database will be available at \url{https://chenfeng-bristol.github.io/MVAD/}.
\end{abstract}

%%%%%%%%% BODY TEXT
\section{Introduction}
\label{sec:intro}

With the significant growth in subscribers to streaming services \cite{Stoll_2023}, such as Netflix and Amazon Prime Video, video streaming applications are now one of the largest consumers of global Internet bandwidth. For video streaming service providers, it is essential to monitor the quality of a user's experience during viewing. However, this process is complex because streamed content can be impacted by multiple visual artifacts introduced at different production and delivery stages including acquisition, post-production, compression, and transmission \cite{bull2021intelligent}. For example, source artifacts such as \textit{motion blur}, \textit{darkness} and \textit{graininess} can be produced during acquisition; \textit{banding} and \textit{aliasing} may be introduced in post-production; video compression can introduce \textit{spatial blur} and \textit{blockiness}; and finally packet loss in transmission leads to \textit{transmission errors}, \textit{frame dropping}, or \textit{black frames}. These artifacts can affect the perceived quality of streamed video content and thus degrade the quality of the user's experience. Therefore, it is important to identify the presence of these artifacts in order to enable appropriate enhancement processes and provide system feedback.

Although video quality assessment is a well-established research area, with numerous classic \cite{ssim,c:mssim, PPVM,niqe,brisque,VIIDEO,VBLIINDS} and learning-based approaches \cite{w:VMAF,  CNNTLVQM, VSFA, VIDEVAL,feng2024rankdvqa} developed over the past two decades, these methods can only provide a quantitative prediction of perceptual quality, rather than detect and identify the artifacts appearing in a video sequence. To achieve accurate artifact detection, some existing works apply thresholding on predicted quality scores to determine the existence of artifacts, with MaxVQA \cite{MaxVQA} being a typical example. Other works have developed bespoke models for individual artifacts, such as CAMBI \cite{tandon2021cambi}, BBAND \cite{tu2020bband}, EFENet \cite{zhao2021defocus}, MLDBD \cite{zhao2023full} and \cite{wolf2008no}. This latter approach is a less practical solution for real-world scenarios where multiple artifacts co-exist and interact. In a recent benchmarking experiment \cite{feng2023bvi}, all these methods were reported to perform poorly on a video database containing multiple source and non-source artifacts - with the average artifact detection accuracy for the best performers being only 55\%.

        \begin{figure*}[htbp]
         \centering
      \centerline{\includegraphics[width=0.89\linewidth]{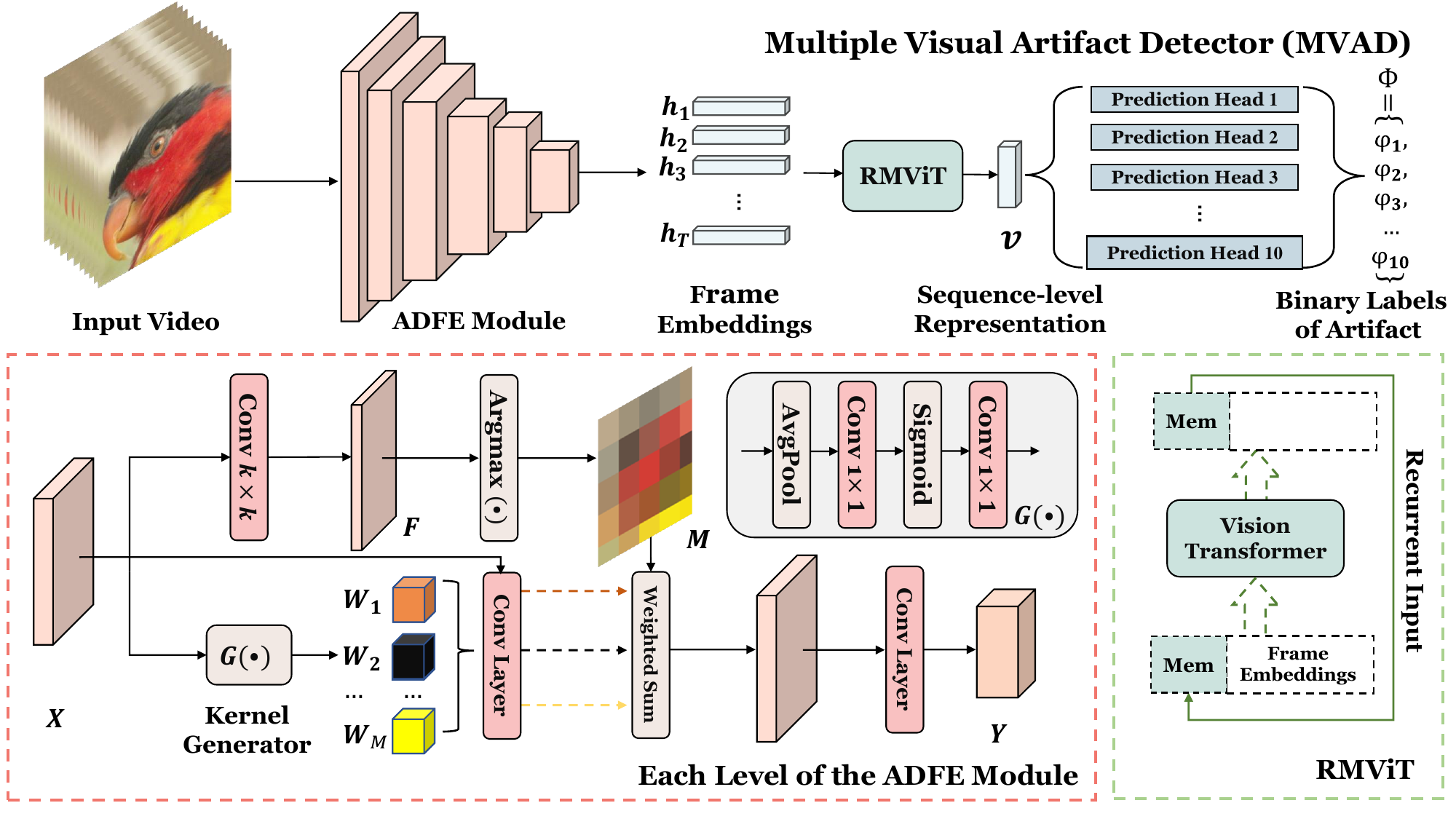}}
             \caption{Illustration of the proposed MVAD framework.}
             \label{fig:framework}
         \end{figure*} 
         
In this context, we propose a new Multiple Visual Artifact Detector, MVAD, which can identify ten different common visual artifacts in streamed video content. As shown in \autoref{fig:framework}, this model employs a novel \textit{Artifact-aware Dynamic Feature Extractor} (ADFE) to extract artifact-relevant spatial features for multiple artifact types, which are then fed into a \textit{Recurrent Memory Vision Transformer} (RMViT) module \cite{bulatov2022recurrent,peng2024rmt} to further capture both global and local temporal information. The obtained spatio-temporal information is then passed to ten \textit{Prediction Heads}, each of which contains a multi-layer perceptron (MLP) to determine a binary label indicating the existence of one type of artifact. To facilitate the optimization of the proposed method, we developed a large and diverse training database based on the Adversarial Data Augmentation strategy \cite{antoniou2017data,volpi2018generalizing}. The primary contributions of this work are summarized as follows:

\begin{itemize}[leftmargin=*]
\item[1)] \textbf{A multi-artifact detection framework:} This work introduces the first multi-artifact detection framework that does not rely on specific video quality assessment models. All the artifact prediction heads in this model are based on the same artifact-aware features, enabling ten artifacts to be detected in a single forward pass of the model, thus significantly improving model efficiency.

\item[2)] \textbf{Artifact-aware Dynamic Feature Extraction}: We designed a new Artifact-aware Dynamic Feature Extractor, which can capture artifact-relevant spatial features (through an end-to-end training process) specifically for all the visual artifacts observed in this model. This is different from existing works \cite{zhao2021defocus,MaxVQA} that employ pre-trained models for feature extraction. It contributes to more precise detection performance by tailoring feature extraction to the artifact detection task. 

\item[3)] \textbf{Training database}: Rather than performing intra-database cross-validation as in existing learning-based artifact detection works \cite{MaxVQA,vidmap}, we generated a large amount of training content through simulating the video streaming pipeline, with each training patch containing up to ten source and non-source artifacts. Through Adversarial Data Augmentation, we created additional more challenging training content in order to enhance the robustness and generalization of the model.

\item[4)] \textbf{Recurrent Memory Vision Transformer (RMViT) module}: The RMViT module demonstrates superior performance in capturing the temporal characteristics of artifacts compared to other pooling methods. This is the first time that the recurrent memory mechanism \cite{bulatov2022recurrent} has been used in the context of artifact detection. It has previously been proved to be effective for long sequence processing when employed in language modeling \cite{bulatov2022recurrent}.
    
\end{itemize}

The proposed MVAD model has been evaluated (with fixed model parameters) on two multi-artifact video databases, Maxwell \cite{MaxVQA} and BVI-Artifact \cite{feng2023bvi}. Results show that MVAD is the best performer in each artifact category, consistently offering superior detection performance compared to  seven other benchmark methods. A comprehensive ablation study has confirmed the effectiveness of all of the primary contributions listed above.
%-------------------------------------------------------------------------
\section{Related Work}
\label{sec:Related Work}

\noindent\textbf{Video quality assessment.} Although subjective quality assessment offers a gold standard for measuring the perceived quality of streamed video content, it is impractical for online delivery, where objective quality models are instead used to predict perceptual video quality in a more efficient manner. Early objective video quality assessment methods \cite{ssim, c:mssim, ssimplus, vssim, MAD, MOVIE, PPVM, niqe, brisque, VIIDEO, VBLIINDS} typically rely on hand-crafted models that are based on classic signal processing theories to exploit different properties of the human visual system. Some of these quality metrics have further been `fused'  with other video features in a regression framework \cite{w:VMAF, STGREED, VIDEVAL} to achieve better prediction performance. More recently, many quality models \cite{Kim2018DeepVQ, C3DVQA, wu2023discovqa, DeepVBQA, VSFA, CNNTLVQM, wang2021rich} have exploited deep network architectures which can learn from subjective data. These  models have been further enhanced based on more advanced training strategies \cite{hou2022perceptual, feng2024rankdvqa, conviqt}, which enables the use of more diverse training material without the need to perform expensive subjective tests.

\noindent\textbf{Artifact Detection.} Existing artifact detection methods can be classified into two groups. The first group focuses on the detection of single artifacts, such as CAMBI \cite{tandon2021cambi} and BBAND \cite{tu2020bband} for banding artifact detection, EFENet \cite{zhao2021defocus} and MLDBD \cite{zhao2023full} for spatial blur detection, and \cite{wolf2008no} for frame dropping. VIDMAP \cite{vidmap} is another notable example which employs nine individual models to identify and quantify the extent of video impairments separately for nine common video artifacts. It is noted that these methods often assume the existence of a single type of artifact in each video, which is not tenable in many practical scenarios where artifacts generated at various stages of video streaming co-exist and interact. A second class of artifact detection method approaches the task from the video quality assessment (VQA) perspective, determining the presence of visual artifacts by thresholding objective quality indices. One of such methods is \cite{MaxVQA}, which can detect eight common artifacts induced during video acquisition and delivery. Existing quality metrics can also be directly used for detecting compression artifacts as in \cite{vidmap} together with static thresholding. However, this has been reported to be less effective compared to specifically designed artifact detectors \cite{vidmap}. 

\noindent\textbf{Artifact databases.} A realistic, diverse  and comprehensive benchmark database is key for evaluating the performance of artifact detectors. As far as we are aware, there are only two databases which are publicly available containing content with multiple artifacts, Maxwell \cite{MaxVQA} and BVI-Artifact \cite{feng2023bvi}. The former consists of 4,543 User-Generated Content (UGC) video sequences at various spatial resolutions, each of which contains up to eight artifacts, while BVI-Artifact includes 480 HD and UHD Professionally-Generated Content (PGC) videos, each with up to six source and non-source artifacts.

\section{Method}
\label{sec:method}

The proposed Multiple Visual Artifact Detector, MVAD, is illustrated in \autoref{fig:framework}. It has been designed to detect multiple pre-defined visual artifacts in a streamed video without the need for a pristine reference. In this work, we focus on ten common visual artifacts in video streaming, as defined in \cite{feng2023bvi}; however this can be re-configured by adding additional \textit{Prediction Heads} according to the requirements of different application scenarios.

In this framework, each frame of the input video signal is first processed by the \textit{Artifact-aware Dynamic Feature Extraction} (ADFE) module, which outputs a frame embedding, $\mathbf{h} \in \mathbb{R}^{2048\times 1}$. All extracted frame embeddings are fed into the \textit{Recurrent Memory Vision Transformer} (RMViT) module to obtain a sequence-level representation $\mathbf{v} \in \mathbb{R}^{128\times 1}$. $\mathbf{v}$ is then shared by ten \textit{Prediction Heads} (with the same network architecture but different model parameters) as input, each of which outputs a binary label $\varphi_j \in \{1, 0\}, j = 1,2\dots \text{ or } 10$ to indicate the presence of an artifact type. The network structures, the training database and the model optimization strategy are described in detail below.

\subsection{Network architecture}

\noindent\textbf{Artifact-aware Dynamic Feature Extraction (ADFE).} Different visual artifacts in streamed content may exhibit distinct spatial and temporal characteristics. For example, \textit{graininess} artifacts are typically uniformly distributed within a video sequence, while \textit{banding} and \textit{motion blur} tend to appear within certain spatial and temporal regions. It is hence challenging to employ a static and pre-trained feature extraction module (as is done in many existing works such as \cite{zhao2021defocus, MaxVQA}) for multi-artifact detection. To address this, we have designed a new Artifact-aware Dynamic Feature Extraction (ADFE) module, inspired by the dynamic region-aware convolution mechanism \cite{chen2021dynamic,li2022contextual,you2023improved} that has been exploited for different high-level vision tasks (image classification, face recognition, object detection, and segmentation~\cite{chen2021dynamic}). Specifically, the ADFE module performs multiple-level feature extraction using a pyramid network. At each level, the input $\mathbf{X}$ (either a single video frame or a feature map processed by the previous feature extraction layer) is first processed by a standard convolutional layer with a kernel size $k\times k$ to produce a region-aware guided feature map, $\mathbf{F}$, which is expected to capture the artifact distribution.  $\mathbf{F}$ is further used to obtain a guided mask $\mathbf{M}$  through $\textrm{argmax}(\cdot)$ operation. The input $\mathbf{X}$ is also fed into a filter generator module $G(\cdot)$ \cite{chen2021dynamic} to produce a series of region-based filters with learnable kernel sizes, $W_1, W_2, \cdots, W_M$, where $M$ is the number of regions (dependent on resolution). All these filters are used by a convolutional layer to process the input $\mathbf{X}$ within each region. The output is weighted by the guided mask $\mathbf{M}$, and then down-sampled by another convolutional layer to obtain the output at this level, $\mathbf{Y}$. In this ADFE module, there are six feature extraction levels employed in total to finally produce a frame embedding, $\mathbf{h}_i \in \mathbb{R}^{2048\times 1}$, where $i$ is the frame index. As the ADFE module is trained with the whole framework in an end-to-end manner, the features here are expected to capture the different global distributions and local dynamics corresponding to all the artifacts observed.

\noindent\textbf{Recurrent Memory Vision Transformer (RMViT). } We employed the same network structure of the RMViT module in \cite{peng2024rmt}, which was inspired by the recurrent memory mechanism \cite{bulatov2022recurrent} that has been used for language models \cite{bulatov2022recurrent, bulatov2023scaling} to handle long sequences. This module is expected to obtain both long-term and short-term temporal information within a video sequence, which has been shown to be effective for the video quality assessment task \cite{peng2024rmt}. The RMViT module consists of multiple recurrent iterations, each of which takes frame embeddings in a sequence segment of length $N=8$, $[\mathbf{h}_{i+1},\mathbf{h}_{i+2},\dots,\mathbf{h}_{i+N}]$, assuming this segment starts from the $i^{th}$ frame of the video, together with memory tokens (either empty ones initially or memory tokens generated in the previous iteration). These tokens are processed by a Vision Transformer \cite{dosovitskiy2020image} producing the output with the same size including renewed memory tokens and current frame embeddings. In the final iteration, the processed memory tokens and the processed frame embeddings in all recurrent iterations are averaged and processed by an MLP layer to generate a sequence level representation $\mathbf{v} \in \mathbb{R}^{128\times 1}$. A more detailed description of RMViT module can be found in \cite{peng2024rmt}. 

\noindent{\textbf{Prediction Heads}.} For each artifact type, MVAD employs an MLP to determine its existence in the given video. Specifically, each \textit{Prediction Head} takes the same video representation $\mathbf{v}$ as input and feeds it into a dropout layer. The output is then passed to a two-layer MLP with GELU activation in the hidden layer and a sigmoid activation at the output to obtain a probability $p$, indicating the probability of the artifact existence:
\begin{equation}
\left\{
\begin{array}{l}
    \varphi=1, \text{ if } p>0.5\\
    \varphi=0, \text{ otherwise } p\leq 0.5
    \end{array}\right..
\end{equation}

\subsection{Training Database}
\label{sec:database}

\begin{figure*}[htbp]
         \centering
         \centerline{\includegraphics[width=0.9\linewidth]{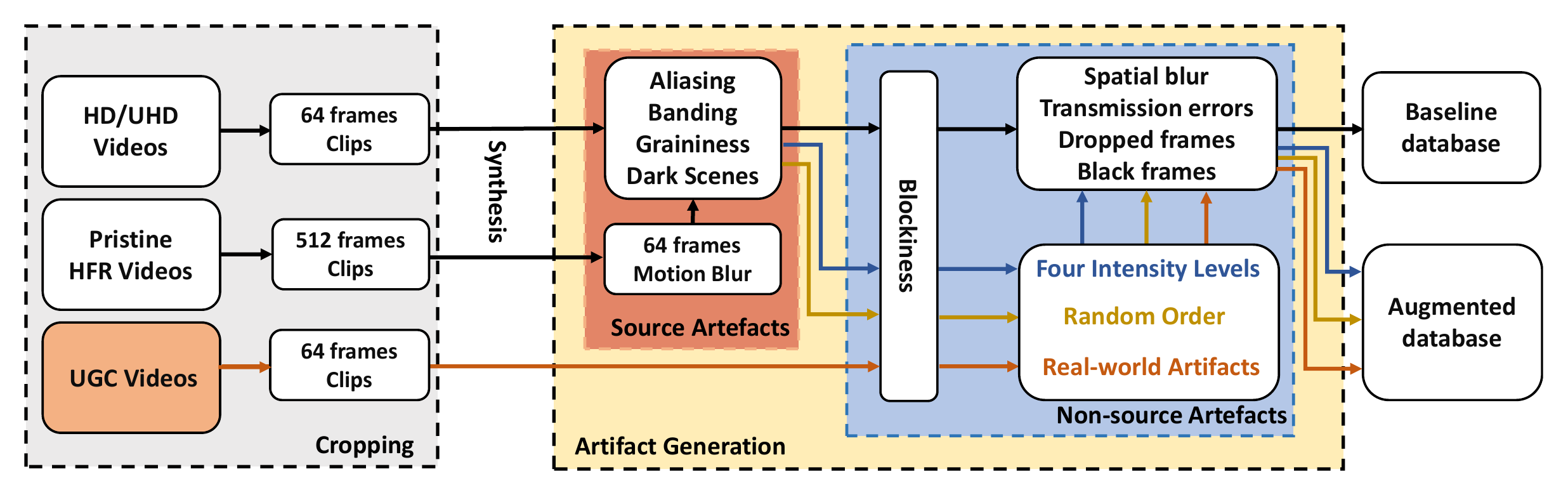}}
             \caption{The workflow used to generate the training material for optimizing the proposed MVAD model.}
             \label{fig:generated}
         \end{figure*}

\noindent\textbf{Baseline database.} To support the training of the proposed MVAD model and enable cross-dataset evaluation, we developed a large and diverse database based on 100 pristine HD/UHD source sequences from NFLX-public \cite{w:VMAF}, BVI-DVC \cite{ma2021bvi} and BVI-CC \cite{katsenou2022bvi} databases, and 100 HFR source videos from LIVE-YT-HFR \cite{madhusudana2021subjective} and Adobe240 \cite{su2017deep}. Based on the collected source content, we followed the workflow illustrated in \autoref{fig:generated}, and randomly cropped each video into six spatio-temporal patches, each with a size of 560 (width) $\times$ 560 (height) $\times$ 64 (length) for HD/UHD content or 560 (width) $\times$ 560 (height) $\times$ 512 (length) for HFR clips. This results in 600 HD/UHD and 600 HFR source patches. 
In this work, we specifically focus on ten visual artifacts that commonly occur in streamed video content, as defined in two existing benchmark databases, MaxVQA \cite{MaxVQA} and BVI-Artifact \cite{feng2023bvi}. These include five source artifacts, \textit{motion blur}, \textit{dark scene} (named \textit{lighting} in Maxwell), \textit{graininess} (\textit{noise}), \textit{aliasing} and \textit{banding}, and five non-source artifacts, \textit{blockiness} (\textit{compression artifacts}), \textit{spatial blur} (\textit{focus}), \textit{transmission errors}, \textit{dropped frames} (\textit{fluency}) and \textit{black frames}. 

For all 600 cropped HFR patches, we first synthesized the motion blur artifact using the approach described in \cite{shen2020blurry}, which reduces the temporal length of these patches from 512 to 64. The resulting 600 patches with \textit{motion blur}, together with those 600 patches cropped from original HD/UHD sources, were further processed to simulate \textit{dark scene, graininess, aliasing} and \textit{banding} artifacts sequentially based on the procedures designed in \cite{vidmap}. Each of these four source artifacts has a 50\% probability of being introduced in every patch mentioned above, which prevents model bias and improves generalization. We repeated the synthesis pipeline four times, and as a result, generated 4800 (1200$\times$4) patches here, each of which contains up to five source artifacts.

For non-source artifact generation, we first used an HEVC codec, x265 \cite{x265} (\textit{medium} preset) to compress each patch. Here we employed two quantization parameter (QP) values, QP32 and QP47. The former emulates a scenario when the video quality is relatively high, while QP47 is used to generate \textit{blockiness} artifacts. For all compressed patches, we further synthesized the other four non-source artifacts, \textit{spatial blur, transmission errors, dropped frames} and \textit{black frames}, in a sequential manner, as shown in \autoref{fig:generated}, based on the synthesis methods described in \cite{vidmap,feng2023bvi}. Again, each of these four non-source artifacts also has a 50\% probability of being synthesized. This process has been repeated four times for each patch, resulting in a total of 38,400 patches (4,800$\times$2$\times$4). 

\noindent\textbf{Augmented database.} To further enhance the robustness and effectiveness of our artifact detection model, following the Adversarial Data Augmentation (ADA) strategy \cite{volpi2018generalizing,antoniou2017data, zhao2020maximum,zhang2019dada}, we generated an augmented dataset containing additional training patches with artifacts at \textit{various intensity levels}, \textit{non-sequentially synthesized artifacts} and \textit{real-world source artifacts}.
   
Based on the 4,800 patches with source artifacts generated for the baseline database, we introduced five non-source artifacts into each patch sequentially as for the baseline database, but at one of four intensity levels (very noticeable, noticeable, subtle, and very subtle, with implementation details in the \textit{Supplementary}). This results in 4,800 additional patches with non-source artifacts at various intensity levels.

To further randomize the order of artifact introduction, based on the 1200 source patches created, we produced more training patches by generating ten types of artifacts in a random sequence. Each artifact, similar to the baseline database, is associated with a 50\% probability of inclusion. This process was repeated four times, producing 4,800 additional training patches with artifacts synthesized in random order.
    
Finally, we collected 60 source sequences from the YouTube-UGC subset \cite{kyncl2017youtube} and randomly cropped six 560$\times$560$\times$64 from each sequence, producing 360 new source patches, which contain real-world source artifacts (rather than synthesized ones) including \textit{motion blur, dark scenes, aliasing}, and \textit{graininess}. They are expected to provide more realistic training samples. Five non-source artifacts were further introduced based on the same methodology used for the baseline dataset, which generated 2,880 (360$\times$2$\times$4) training patches.

In total, the two training datasets contain 50,880 (all in 560$\times$560$\times$64 size) training patches. Each of these has been annotated with binary artifact labels (i.e. ten labels per patch video, corresponding to ten artifacts) to support the supervised learning process.

\subsection{Training Strategy}

The proposed network architecture has been trained from scratch in an end-to-end manner, using a combination of contrastive loss \cite{khosla2020supervised} and binary cross-entropy loss. Specifically, for each batch with randomly selected $B$ training patches, the overall loss $\mathcal{L}$ is calculated as a weighted sum of the contrastive loss and the binary cross-entropy loss for all $B$ patches in a batch:
\begin{equation}
    \mathcal{L} = \sum_{i=1}^{B} \left(\alpha \mathcal{L}^i_\text{contrastive} + \beta \mathcal{L}^i_\mathrm{BCE}\right),
\end{equation}
where $\alpha=0.5$ and $\beta=0.5$ are the weights to trade off the relationship between the contrastive loss and the BCE loss, respectively. The contrastive loss $\mathcal{L}^i_\text{contrastive}$ for the $i^{th}$ patch is defined as \cite{MaxVQA}:
% \begin{equation}
% \displaystyle
%     \mathcal{L}^i_\text{contrastive} = -\frac{1}{B} \frac{\sum_{j=1}^B (\frac{1}{ \tau}\mathrm{sim}(\mathbf{v}_i,\mathbf{v}_j) (\mathbf{\Phi_i} \cdot \mathbf{\Phi_j}))}{\sum_{j=1}^B (\mathbf{\Phi_i} \cdot \mathbf{\Phi_j})}\log \left( \sum_{k=1,k\neq i}^B \exp(\frac{1}{ \tau}\mathrm{sim}(\mathbf{v}_i,\mathbf{v}_k)) \right),
% \end{equation}
\begin{equation}
\displaystyle
    \mathcal{L}^i_\text{contrastive} = \frac{1}{|P(i)|}\sum_{j\in P(i)}-\log \frac{\exp(\frac{1}{ \tau}\mathrm{sim}(\mathbf{v}_i,\mathbf{v}_j)) }{ \sum_{k=1,k\neq i}^B \exp(\frac{1}{ \tau}\mathrm{sim}(\mathbf{v}_i,\mathbf{v}_k)) )},
\end{equation}
in which $\tau$ is the temperature parameter that scales the cosine similarity.  $\mathbf{v}_i$ is the sequence-level representation for the $i^{th}$ patch, and $P(i)$ denotes the set of indices of the patches with the same artifact labels (positive pairs) in the same batch as the $i^{th}$ patch. The symbol $\cdot$ represents the dot product between vectors. The function $\text{sim}(\cdot)$ stands for the cosine similarity between two video representations.
\begin{equation}
    \text{sim}({\mathbf{v}_i,\mathbf{v}_j}) = \frac{\mathbf{v}_i \cdot \mathbf{v}_j}{\|\mathbf{v}_i\| \|\mathbf{v}_j\|},
\end{equation}
Here the term $\|\mathbf{v}_i\|$  and $\|\mathbf{v}_j\|$ represents the L2-norm (vector magnitude) of $\mathbf{v}_i$ and $\mathbf{v}_j$, respectively.

In addition, as for many other standard classification tasks \cite{rawat2017deep,zhang2018generalized, ruby2020binary}, we use the binary cross-entropy loss $\mathcal{L}_\mathrm{BCE}$ to match the output of the \textit{Prediction Heads} to ten ground-truth binary labels.  The BCE loss for the $i^{th}$ patch, $\mathcal{L}_\mathrm{BCE}^i$, is given below:
\begin{equation}
    \mathcal{L}_\mathrm{BCE}^i = -\frac{1}{10} \sum_{j=1}^{10} \left[ \boldsymbol{\psi}_i^j \log (p_i^j) + (1 - \boldsymbol{\psi}_i^j) \log (1 - p_i^j) \right],
\end{equation}
where $\boldsymbol{\psi}_i^j$ stands for the ground truth binary label for artifact $j$ in the $i^{th}$ patch, and $p_i^j$ is the predicted binary label for artifact $j$ in the $i^{th}$ patch.

\section{Experiment Configuration}
\label{sec:Experiments}

\begin{table*}[t!]
    \centering
    \resizebox{0.9\linewidth}{!}{\begin{tabular}{c r c c c c c c c c c c}
    \toprule
       Metric  & Method  & Motion & Dark (Lighting) & Grain.(Noise) & Block. (Compression)& Spat.(Focus) & Drop. (Fluency)\\
       \midrule
      \multirow{6}{*}{Acc. (\%) $\uparrow$} 
       & EFENet \cite{zhao2021defocus} &\cellcolor[HTML]{EFEFEF}- &\cellcolor[HTML]{EFEFEF}- &\cellcolor[HTML]{EFEFEF}- &\cellcolor[HTML]{EFEFEF}- & 54.55 &\cellcolor[HTML]{EFEFEF}- \\
      & MLDBD \cite{zhao2023full} &\cellcolor[HTML]{EFEFEF}- &\cellcolor[HTML]{EFEFEF}- &\cellcolor[HTML]{EFEFEF}- &\cellcolor[HTML]{EFEFEF}- & 56.38 &\cellcolor[HTML]{EFEFEF}- \\
      & Wolf \textit{et al.} \cite{wolf2008no} &\cellcolor[HTML]{EFEFEF}- &\cellcolor[HTML]{EFEFEF}- &\cellcolor[HTML]{EFEFEF}- &\cellcolor[HTML]{EFEFEF}- &\cellcolor[HTML]{EFEFEF}- & 50.63 \\
      & VIDMAP \cite{vidmap} &\cellcolor[HTML]{EFEFEF}- &\cellcolor[HTML]{EFEFEF}- &\cellcolor[HTML]{EFEFEF}- & 70.90 & 58.96 & 59.20 \\
      & MaxVQA \cite{MaxVQA} & 78.30 & 75.69 & 65.10 & 85.42 & 82.66 & 79.66 \\
      & \textbf{MVAD (ours)} & \textbf{82.64} & \textbf{79.91} & \textbf{72.67} & \textbf{98.00} & \textbf{90.46} & \textbf{81.68} \\
      \midrule
      \multirow{6}{*}{F1 $\uparrow$} 
      & EFENet \cite{zhao2021defocus} &\cellcolor[HTML]{EFEFEF}- &\cellcolor[HTML]{EFEFEF}- &\cellcolor[HTML]{EFEFEF}- &\cellcolor[HTML]{EFEFEF}- & 0.66 &\cellcolor[HTML]{EFEFEF}- \\
      & MLDBD \cite{zhao2023full} &\cellcolor[HTML]{EFEFEF}- &\cellcolor[HTML]{EFEFEF}- &\cellcolor[HTML]{EFEFEF}- &\cellcolor[HTML]{EFEFEF}- & 0.63 &\cellcolor[HTML]{EFEFEF}- \\
      & Wolf \textit{et al.} \cite{wolf2008no} &\cellcolor[HTML]{EFEFEF}- &\cellcolor[HTML]{EFEFEF}- &\cellcolor[HTML]{EFEFEF}- &\cellcolor[HTML]{EFEFEF}- &\cellcolor[HTML]{EFEFEF}- & 0.59 \\
      & VIDMAP \cite{vidmap} &\cellcolor[HTML]{EFEFEF}- &\cellcolor[HTML]{EFEFEF}- &\cellcolor[HTML]{EFEFEF}- & 0.71 & 0.64 & 0.62 \\
      & MaxVQA \cite{MaxVQA} & 0.75 & 0.72 & 0.66 & 0.85 & 0.82 & 0.77 \\
      & \textbf{MVAD (ours)} & \textbf{0.78} & \textbf{0.80} & \textbf{0.72} & \textbf{0.99} & \textbf{0.92} & \textbf{0.79} \\
      \midrule
      \multirow{6}{*}{AUC $\uparrow$} 
      & EFENet \cite{zhao2021defocus} &\cellcolor[HTML]{EFEFEF}- &\cellcolor[HTML]{EFEFEF}- &\cellcolor[HTML]{EFEFEF}- &\cellcolor[HTML]{EFEFEF}- & 0.63 &\cellcolor[HTML]{EFEFEF}- \\
      & MLDBD \cite{zhao2023full} &\cellcolor[HTML]{EFEFEF}- &\cellcolor[HTML]{EFEFEF}- &\cellcolor[HTML]{EFEFEF}- &\cellcolor[HTML]{EFEFEF}- & 0.61 &\cellcolor[HTML]{EFEFEF}- \\
      & Wolf \textit{et al.} \cite{wolf2008no} &\cellcolor[HTML]{EFEFEF}- &\cellcolor[HTML]{EFEFEF}- &\cellcolor[HTML]{EFEFEF}- &\cellcolor[HTML]{EFEFEF}- &\cellcolor[HTML]{EFEFEF}- & 0.62 \\
      & VIDMAP \cite{vidmap} &\cellcolor[HTML]{EFEFEF}- &\cellcolor[HTML]{EFEFEF}- &\cellcolor[HTML]{EFEFEF}- & 0.70 & 0.62 & 0.64 \\
      & MaxVQA \cite{MaxVQA} & 0.81 & 0.76 & 0.63 & 0.88 & 0.87 & \textbf{0.78} \\
      & \textbf{MVAD (ours)} & \textbf{0.85} & \textbf{0.78} & \textbf{0.68} & \textbf{0.99} & \textbf{0.90} & \textbf{0.78} \\
      \bottomrule
    \end{tabular}}
        \caption{Artifact detection results on the Maxwell database \cite{MaxVQA}. Here `\colorbox{mygray}{ - }' indicates that the tested method in this row is not designed to identify the corresponding artifact in this column.\label{tab:maxvqa}}
\end{table*}

\noindent\textbf{Implementation Details.} Pytorch 1.12 was used to implement the proposed network architecture, with the following training parameters: ADAM optimization \cite{kingma2014adam} with parameter settings $\beta_1$=0.9 and $\beta_2$=0.999; 50 training epochs; batch size of 8; the initial learning rate is 0.001 with weight decay of 0.05 after 10 epochs. Temperature parameter $\tau$ is set to 0.1. Kernel size $k=3$. This experiment was executed on a computer with a 2.4GHz Intel CPU and an NVIDIA 3090 GPU.

\noindent\textbf{Evaluation settings.} To evaluate the performance of the proposed method, we have employed two public artifact databases, Maxwell \cite{MaxVQA} and BVI-Artifact \cite{feng2023bvi} in the benchmark experiment. As described in \autoref{sec:Related Work}, Maxwell contains UGC videos with eight artifacts, while BVI-Artifact consists of PGC content associated with ten artifacts (as we defined in this work). Both test datasets do not contain sequences which are included in the training database. To test model generalization, we did not perform intra-database cross validation for all tested artifact detectors. Instead, we fixed all the optimized model parameters in the inference phase or used the pre-trained models for benchmark methods. It is noted that the Maxwell \cite{MaxVQA} database contains content with eight artifacts, but we only test the six of them that are relevant to video streaming, including \textit{motion blur, dark scene (lighting), graininess (noise), blockiness (compression artifacts), spatial blur (focus)} and \textit{droppped frames (fluency)}\footnote{The Maxwell database employs a threshold on the collected opinion scores (in different dimensions) to determine if a test sequence contains certain artifacts. Here we followed this practice to obtain binary labels as ground truth in this experiment.}.

We have benchmarked the proposed MVAD model against seven existing artifact detectors, among which MaxVQA \cite{MaxVQA} and VIDMAP \cite{vidmap} can detector multiple artifacts, while CAMBI \cite{tandon2021cambi}, BBAND \cite{tu2020bband}, EFENet \cite{zhao2021defocus}, MLDBD \cite{zhao2023full} and Wolf \textit{el at.} \cite{wolf2008no} are single artifact detectors. Their implementations are based on their original literature and the practice in \cite{MaxVQA,feng2023bvi}. For the \textit{black screen} artifact in the BVI-Artifact database, as we did not find benchmark methods, we solely provide the results for the proposed model.

To measure the artifact detection performance, three commonly used metrics are employed here including detection accuracy (Acc.), F1 score \cite{vidmap}, and the AUC (area under curve) index. AUC values are calculated through changing the default detection threshold in each method and drawing the receiver operating characteristic (ROC) curves as in \cite{feng2023bvi}. 

\section{Results and Discussion}
 
\subsection{Overall performance}
         
\autoref{tab:maxvqa} and \autoref{tab:bviartifact} summarize the detection performance of the proposed method compared to the other seven artifact detectors on two test databases. It can be observed that MVAD outperforms all the other methods in each artifact category across the two databases. The detection performance is particularly superior (above 90\%) for \textit{blockiness, spatial blur} and \textit{aliasing} artifacts, much higher than that of the second best performers. However it is also clear, despite the evident performance improvements, that the proposed method can be further enhanced for challenging artifact cases such as \textit{motion blur, graininess} and \textit{banding}. We have plotted the ROC figures for each artifact category in two databases and provided in \textit{Supplementary}, which also confirms the effectiveness of the proposed method from a different perspective.

\begin{table*}[t!]
    \centering

    \resizebox{0.9\linewidth}{!}{\begin{tabular}{c r c c c c c c c c c c}
    \toprule
       Metric  & Method  & Motion & Dark & Grain. & Alias. & Band. & Block. & Spat. & Drop. & Trans. & Black.\\
       \midrule
      \multirow{8}{*}{Acc. (\%) $\uparrow$} 
      & CAMBI \cite{tandon2021cambi} & \cellcolor[HTML]{EFEFEF}- & \cellcolor[HTML]{EFEFEF}- & \cellcolor[HTML]{EFEFEF}- & \cellcolor[HTML]{EFEFEF}- & 61.88 & \cellcolor[HTML]{EFEFEF}- & \cellcolor[HTML]{EFEFEF}- & \cellcolor[HTML]{EFEFEF}- & \cellcolor[HTML]{EFEFEF}- & \cellcolor[HTML]{EFEFEF}- \\
      & BBAND \cite{tu2020bband} & \cellcolor[HTML]{EFEFEF}- & \cellcolor[HTML]{EFEFEF}- & \cellcolor[HTML]{EFEFEF}- & \cellcolor[HTML]{EFEFEF}- & 50.00 & \cellcolor[HTML]{EFEFEF}- & \cellcolor[HTML]{EFEFEF}- & \cellcolor[HTML]{EFEFEF}- & \cellcolor[HTML]{EFEFEF}- & \cellcolor[HTML]{EFEFEF}- \\
      & EFENet \cite{zhao2021defocus} & \cellcolor[HTML]{EFEFEF}- & \cellcolor[HTML]{EFEFEF}- & \cellcolor[HTML]{EFEFEF}- & \cellcolor[HTML]{EFEFEF}- & \cellcolor[HTML]{EFEFEF}- & \cellcolor[HTML]{EFEFEF}- & 47.08 & \cellcolor[HTML]{EFEFEF}- & \cellcolor[HTML]{EFEFEF}- & \cellcolor[HTML]{EFEFEF}- \\
      & MLDBD \cite{zhao2023full} & \cellcolor[HTML]{EFEFEF}- & \cellcolor[HTML]{EFEFEF}- & \cellcolor[HTML]{EFEFEF}- & \cellcolor[HTML]{EFEFEF}- & \cellcolor[HTML]{EFEFEF}- & \cellcolor[HTML]{EFEFEF}- & 49.58 & \cellcolor[HTML]{EFEFEF}- & \cellcolor[HTML]{EFEFEF}- & \cellcolor[HTML]{EFEFEF}- \\
      & Wolf \textit{et al.} \cite{wolf2008no} & \cellcolor[HTML]{EFEFEF}- & \cellcolor[HTML]{EFEFEF}- & \cellcolor[HTML]{EFEFEF}- & \cellcolor[HTML]{EFEFEF}- & \cellcolor[HTML]{EFEFEF}- & \cellcolor[HTML]{EFEFEF}- & \cellcolor[HTML]{EFEFEF}- & 51.67 & \cellcolor[HTML]{EFEFEF}- & \cellcolor[HTML]{EFEFEF}- \\
      & VIDMAP \cite{vidmap} & \cellcolor[HTML]{EFEFEF}- & \cellcolor[HTML]{EFEFEF}- & \cellcolor[HTML]{EFEFEF}- & 50.00 & 56.25 & 54.38 & 47.29 & 45.42 & 51.04 & \cellcolor[HTML]{EFEFEF}- \\
      & MaxVQA \cite{MaxVQA} & 51.88 & 73.13 & 38.75 & \cellcolor[HTML]{EFEFEF}- & \cellcolor[HTML]{EFEFEF}- & 64.58 & 53.54 & \cellcolor[HTML]{EFEFEF}- & \cellcolor[HTML]{EFEFEF}- & \cellcolor[HTML]{EFEFEF}- \\
      & \textbf{MVAD (ours)} & \textbf{59.38} & \textbf{78.32} & \textbf{66.87} & \textbf{93.75} & \textbf{64.96} & \textbf{99.97} & \textbf{92.92} & \textbf{71.25} & \textbf{74.17} & \textbf{76.25} \\
      \midrule
      \multirow{8}{*}{F1 $\uparrow$} 
      & CAMBI \cite{tandon2021cambi} & \cellcolor[HTML]{EFEFEF}- & \cellcolor[HTML]{EFEFEF}- & \cellcolor[HTML]{EFEFEF}- & \cellcolor[HTML]{EFEFEF}- & 0.53 & \cellcolor[HTML]{EFEFEF}- & \cellcolor[HTML]{EFEFEF}- & \cellcolor[HTML]{EFEFEF}- & \cellcolor[HTML]{EFEFEF}- & \cellcolor[HTML]{EFEFEF}- \\
      & BBAND \cite{tu2020bband} & \cellcolor[HTML]{EFEFEF}- & \cellcolor[HTML]{EFEFEF}- & \cellcolor[HTML]{EFEFEF}- & \cellcolor[HTML]{EFEFEF}- & 0.44 & \cellcolor[HTML]{EFEFEF}- & \cellcolor[HTML]{EFEFEF}- & \cellcolor[HTML]{EFEFEF}- & \cellcolor[HTML]{EFEFEF}- & \cellcolor[HTML]{EFEFEF}- \\
      & EFENet \cite{zhao2021defocus} & \cellcolor[HTML]{EFEFEF}- & \cellcolor[HTML]{EFEFEF}- & \cellcolor[HTML]{EFEFEF}- & \cellcolor[HTML]{EFEFEF}- & \cellcolor[HTML]{EFEFEF}- & \cellcolor[HTML]{EFEFEF}- & 0.64 & \cellcolor[HTML]{EFEFEF}- & \cellcolor[HTML]{EFEFEF}- & \cellcolor[HTML]{EFEFEF}- \\
      & MLDBD \cite{zhao2023full} & \cellcolor[HTML]{EFEFEF}- & \cellcolor[HTML]{EFEFEF}- & \cellcolor[HTML]{EFEFEF}- & \cellcolor[HTML]{EFEFEF}- & \cellcolor[HTML]{EFEFEF}- & \cellcolor[HTML]{EFEFEF}- & 0.65 & \cellcolor[HTML]{EFEFEF}- & \cellcolor[HTML]{EFEFEF}- & \cellcolor[HTML]{EFEFEF}- \\
      & Wolf \textit{et al.} \cite{wolf2008no} & \cellcolor[HTML]{EFEFEF}- & \cellcolor[HTML]{EFEFEF}- & \cellcolor[HTML]{EFEFEF}- & \cellcolor[HTML]{EFEFEF}- & \cellcolor[HTML]{EFEFEF}- & \cellcolor[HTML]{EFEFEF}- & \cellcolor[HTML]{EFEFEF}- & 0.18 & \cellcolor[HTML]{EFEFEF}- & \cellcolor[HTML]{EFEFEF}- \\
      & VIDMAP \cite{vidmap} & \cellcolor[HTML]{EFEFEF}- & \cellcolor[HTML]{EFEFEF}- & \cellcolor[HTML]{EFEFEF}- & 0.67 & 0.59 & 0.69 & 0.64 & 0.59 & 0.65 & \cellcolor[HTML]{EFEFEF}- \\
      & MaxVQA \cite{MaxVQA} & 0.68 & 0.67 & 0.16 & \cellcolor[HTML]{EFEFEF}- & \cellcolor[HTML]{EFEFEF}- & 0.55 & 0.40 & \cellcolor[HTML]{EFEFEF}- & \cellcolor[HTML]{EFEFEF}- & \cellcolor[HTML]{EFEFEF}- \\
      & \textbf{MVAD (ours)} & \textbf{0.69} & \textbf{0.73} & \textbf{0.46} & \textbf{0.90} & \textbf{0.64} & \textbf{0.99} & \textbf{0.90} & \textbf{0.65} & \textbf{0.73} & \textbf{0.65} \\
      \midrule
      \multirow{8}{*}{AUC $\uparrow$} 
      & CAMBI \cite{tandon2021cambi} & \cellcolor[HTML]{EFEFEF}- & \cellcolor[HTML]{EFEFEF}- & \cellcolor[HTML]{EFEFEF}- & \cellcolor[HTML]{EFEFEF}- & 0.63 & \cellcolor[HTML]{EFEFEF}- & \cellcolor[HTML]{EFEFEF}- & \cellcolor[HTML]{EFEFEF}- & \cellcolor[HTML]{EFEFEF}- & \cellcolor[HTML]{EFEFEF}- \\
      & BBAND \cite{tu2020bband} & \cellcolor[HTML]{EFEFEF}- & \cellcolor[HTML]{EFEFEF}- & \cellcolor[HTML]{EFEFEF}- & \cellcolor[HTML]{EFEFEF}- & 0.51 & \cellcolor[HTML]{EFEFEF}- & \cellcolor[HTML]{EFEFEF}- & \cellcolor[HTML]{EFEFEF}- & \cellcolor[HTML]{EFEFEF}- & \cellcolor[HTML]{EFEFEF}- \\
      & EFENet \cite{zhao2021defocus} & \cellcolor[HTML]{EFEFEF}- & \cellcolor[HTML]{EFEFEF}- & \cellcolor[HTML]{EFEFEF}- & \cellcolor[HTML]{EFEFEF}- & \cellcolor[HTML]{EFEFEF}- & \cellcolor[HTML]{EFEFEF}- & 0.57 & \cellcolor[HTML]{EFEFEF}- & \cellcolor[HTML]{EFEFEF}- & \cellcolor[HTML]{EFEFEF}- \\
      & MLDBD \cite{zhao2023full} & \cellcolor[HTML]{EFEFEF}- & \cellcolor[HTML]{EFEFEF}- & \cellcolor[HTML]{EFEFEF}- & \cellcolor[HTML]{EFEFEF}- & \cellcolor[HTML]{EFEFEF}- & \cellcolor[HTML]{EFEFEF}- & 0.53 & \cellcolor[HTML]{EFEFEF}- & \cellcolor[HTML]{EFEFEF}- & \cellcolor[HTML]{EFEFEF}- \\
      & Wolf \textit{et al.} \cite{wolf2008no} & \cellcolor[HTML]{EFEFEF}- & \cellcolor[HTML]{EFEFEF}- & \cellcolor[HTML]{EFEFEF}- & \cellcolor[HTML]{EFEFEF}- & \cellcolor[HTML]{EFEFEF}- & \cellcolor[HTML]{EFEFEF}- & \cellcolor[HTML]{EFEFEF}- & 0.60 & \cellcolor[HTML]{EFEFEF}- & \cellcolor[HTML]{EFEFEF}- \\
      & VIDMAP \cite{vidmap} & \cellcolor[HTML]{EFEFEF}- & \cellcolor[HTML]{EFEFEF}- & \cellcolor[HTML]{EFEFEF}- & 0.58 & 0.58 & 0.61 & 0.38 & 0.47 & 0.50 & \cellcolor[HTML]{EFEFEF}- \\
      & MaxVQA \cite{MaxVQA} & 0.56 & \textbf{0.84} & 0.36 & \cellcolor[HTML]{EFEFEF}- & \cellcolor[HTML]{EFEFEF}- & 0.80 & 0.54 & \cellcolor[HTML]{EFEFEF}- & \cellcolor[HTML]{EFEFEF}- & \cellcolor[HTML]{EFEFEF}- \\
      & \textbf{MVAD (ours)} & \textbf{0.60} & 0.78 & \textbf{0.56} & \textbf{0.93} & \textbf{0.67} & \textbf{0.99} & \textbf{0.93} & \textbf{0.71} & \textbf{0.80} & \textbf{0.50} \\
      \bottomrule
    \end{tabular}}
    \caption{Artifact detection results on the BVI-Artifact database \cite{feng2023bvi}. Here `\colorbox{mygray}{ - }' indicates that the tested method in this row is not designed to identify the corresponding artifact in this column. \label{tab:bviartifact}}
\end{table*}

 \begin{figure*}[t!]
         \centering
         \centerline{\includegraphics[width=0.9\linewidth]{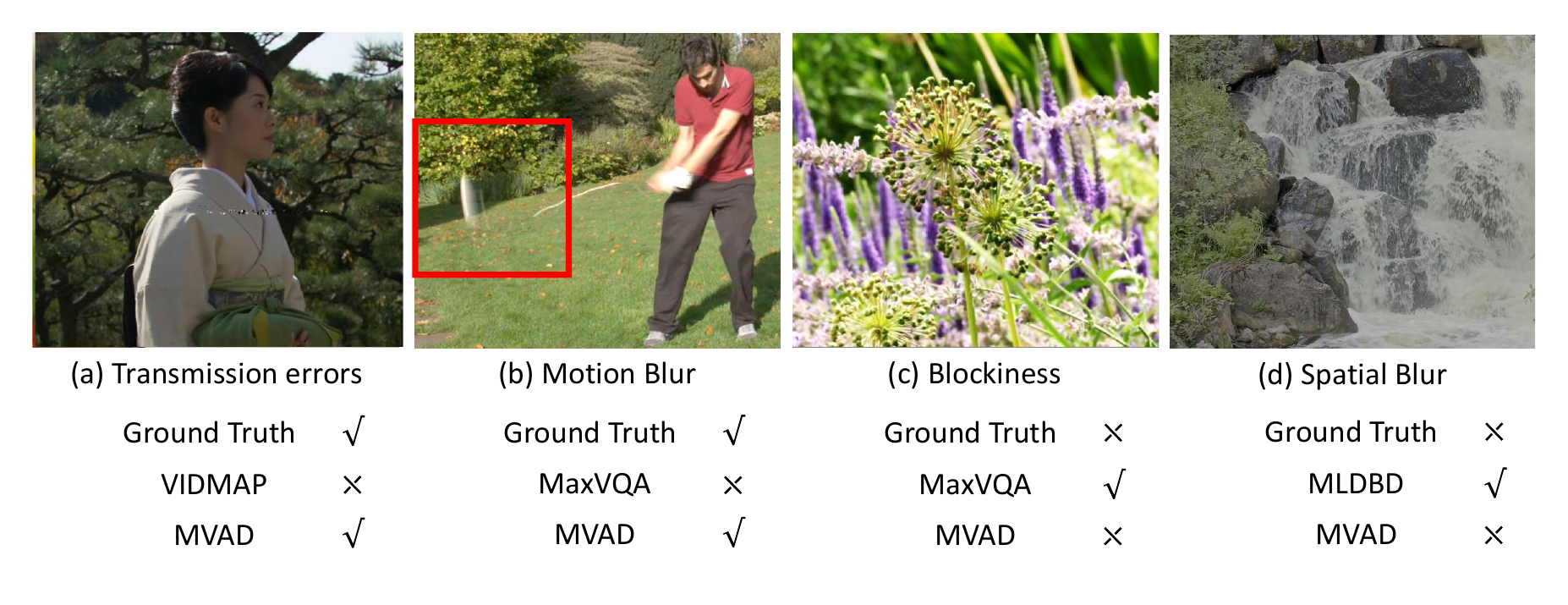}}
             \caption{Visual comparison results showing the effectiveness of the proposed method. In these cases, MVAD offers correct prediction results as ground truth labels, while the comparison methods fail to do so.}
             \label{fig:visual}
         \end{figure*}

In addition, we provide visual qualitative comparisons in \autoref{fig:visual}. In these cases, our approach offered correct prediction results while other existing methods did not.

\subsection{Ablation Study}
\label{sec:ablation styduy}

\begin{table*}[t!]
    \centering
     
    \resizebox{0.9\linewidth}{!}{\begin{tabular}{c c c c c c c c c c c c}
    \toprule
       Metric  & Method  & Motion & Dark & Grain. & Alias. & Band. & Block. & Spat. & Drop. & Trans. & Black.\\
       \midrule
      \multirow{4}{*}{Acc. (\%) $\uparrow$} & v1 & 51.25 & 56.25 & 64.37 & 81.25 & 55.0 & 99.58 & 91.04 & \textbf{71.25} & 60.50 & 75.88\\
      &v2 & 56.25 & 50.25 & 54.37 & 62.50 & 50.0 & 90.63 & 90.0 & 64.58 & 68.75 & 59.58\\
      &v3 & \textbf{59.38} & 74.50 & 66.25 & 90.0 & 62.25 & 99.58 & 90.0 & 60.67 & 58.37 & 50.25\\
      &v4 & 58.58 & 78.25 & \textbf{66.87} & \textbf{93.75} & 62.25 & \textbf{99.97} & 91.75 & 68.42 & 62.50 & 52.46\\
      &\textbf{MVAD}& \textbf{59.38} & \textbf{78.32} & \textbf{66.87} & \textbf{93.75} & \textbf{64.96} & \textbf{99.97} & \textbf{92.92} & \textbf{71.25} & \textbf{74.17} & \textbf{76.25} \\
      \midrule
      \multirow{4}{*}{F1 $\uparrow$} & v1& 0.56 & 0.60 & \textbf{0.46} & 0.85 & 0.58 & \textbf{0.99} & \textbf{0.90} & 0.64 & 0.65 & \textbf{0.65} \\
      &v2 & 0.59 & 0.65 & 0.40 & 0.60 & 0.42 & 0.90 & 0.86 & 0.48 & 0.69 & 0.62 \\
      &v3 & \textbf{0.69} & 0.70 & 0.42 & \textbf{0.90} & 0.62 & \textbf{0.99} & 0.89 & 0.52 & 0.52 & 0.46 \\
      &v4 & 0.58 & \textbf{0.73} & \textbf{0.46} & \textbf{0.90} & 0.58 & \textbf{0.99} & \textbf{0.90} & 0.60 & 0.58 & 0.52 \\
      &\textbf{MVAD} & \textbf{0.69} & \textbf{0.73} & \textbf{0.46} & \textbf{0.90} & \textbf{0.64} & \textbf{0.99} & \textbf{0.90} & \textbf{0.65} & \textbf{0.73} & \textbf{0.65} \\
      \midrule
      \multirow{4}{*}{AUC $\uparrow$} & v1 & 0.51 & 0.61 & 0.54 & 0.86 & 0.56 & \textbf{0.99} & \textbf{0.93} & \textbf{0.71} & 0.56 & \textbf{0.52} \\
      &v2 & 0.56 & 0.70 & 0.52 & 0.58 & 0.51 & 0.93 & 0.90 & 0.57 & 0.74 & 0.48 \\
      &v3 & \textbf{0.60} & 0.75 & 0.52 & 0.82 & 0.61 & \textbf{0.99} & \textbf{0.93} & 0.48 & 0.54 & 0.39 \\
      &v4 & 0.51 & 0.75 & 0.53 & 0.92 & 0.55 & \textbf{0.99} & \textbf{0.93} & 0.68 & 0.48 & 0.45 \\
      &\textbf{MVAD}& \textbf{0.60} & \textbf{0.78} & \textbf{0.56} & \textbf{0.93} & \textbf{0.67} & \textbf{0.99} & \textbf{0.93} & \textbf{0.71} & \textbf{0.80} & {0.50} \\
      \bottomrule
    \end{tabular}}
    \caption{Ablation study results based on the BVI-Artifact database \cite{feng2023bvi}.\label{tab:ablation}}
\end{table*}

\begin{table*}[t!]
    \centering
    \resizebox{0.85\textwidth}{!}{\begin{tabular}{r|c c c c c c c c}
    \toprule
        Complexity&  MaxVQA & VIDMAP & CAMBI& BBAND  & EFENet & MLDBD  & Wolf \textit{et al.}  & \textbf{MVAD}\\\midrule 
         Runtime (s) & 78.64& 1538.51 & 647.38 & 804.74 &  648.19 & 5940.48 & 58.37 & 147.74\\
         Model Size (MB) & 47.6 & 2.8 &\cellcolor[HTML]{EFEFEF}-  &\cellcolor[HTML]{EFEFEF}- & 36.5& 1043.9 &\cellcolor[HTML]{EFEFEF}- & 653.4\\
    \bottomrule 
    
    \end{tabular}}
        \caption{Complexity figures of all eight artifact detectors. `\colorbox{mygray}{ - }' indicates non-deep learning based methods.\label{tab:complexity}}
            
\end{table*}

In order to verify the effectiveness of the primary contributions described in \autoref{sec:intro}, we performed ablation studies to generate four variants of MVAD and compared their performance with the original model, results shown in \autoref{tab:ablation}. Only the BVI-Artifact database was used in this study (as it contains all ten artifact types tested). 

\textbf{Artifact-aware Dynamic Feature Extraction (ADFE).} To confirm the contribution of the proposed ADFE module, we replaced it with the feature extractor used in \cite{feng2024rankdvqa,hou2022perceptual} (a pyramid network without the aritifact-aware masking) and obtained (v1). It can be observed that the performance of (v1) is worse than the original MVAD for most artifact classes, which verifies the importance of the proposed ADFE module. In the \textit{Supplementary}, we have provided a visualization example of the guided mask to demonstrate its influence.

\textbf{Training database.} As a suitable training database to replace the one developed in this work is not available, we could not directly verify its contribution. Instead, we assessed the effectiveness of Adversarial Data Augmentation in enhancing model robustness and generalization. Here we trained the same network architecture as MVAD but only using the baseline training dataset mentioned in \autoref{sec:database} for model optimization, and generated variant (v2). When comparing the performance of (v2) and full MVAD, the performance improvement is evident, in particular for source artifacts such as \textit{dark scene}, \textit{aliasing} and \textit{banding}.

\textbf{RMViT.} Although the effectiveness of RMViT has already been confirmed for large language models \cite{bulatov2022recurrent,bulatov2023scaling} and the video quality assessment task \cite{peng2024rmt}, its contribution to artifact detection is unknown. Here we replaced RMViT with the Gate Recurrent Unit (GRU) \cite{cho2014properties}, which has been used in \cite{conviqt} for video quality assessment. This results in (v3). We further replaced RMViT with simple average pooling and obtained (v4). Based on the results in \autoref{tab:ablation}, we can observe that both (v3) and (v4) are outperformed by MVAD, which proves the effectiveness of the RMViT module.

\subsection{Complexity Analysis}
\label{subsec:complexity}
The complexity figures of the proposed method and seven benchmark approaches are provided in \autoref{tab:complexity}. It is noted that MVAD, MaxVQA and VIDMAP can detect multiple visual artifacts; CAMBI, BBAND, EFENet and MLDBD and Wolf \textit{et al.} are single artifact detectors. Among these models, MVAD has a relatively large model size and slow runtime value compared to MaxVQA.

\section{Conclusion}
\label{sec:conclusion}
 In this paper, we proposed a novel Multiple Visual Artifact Detector (MVAD) for video streaming. It comprises a new Artifact-aware Dynamic Feature Extractor (ADFE) and a Recurrent Memory Vision Transformer (RMViT) module to capture spatial and temporal information for multiple Prediction Heads. It outputs binary artifact predictions for the presence of ten common visual artifacts. We also developed a large and diverse training dataset based on Adversarial Data Augmentation to optimize the proposed model. This multi-artifact detector, MVAD, is the first of its type that does not rely on video quality assessment models. We demonstrated its superior performance compared to seven existing artifact detection methods on two large benchmark databases. 
 
 Future work should focus on reducing model complexity using knowledge distillation \cite{hinton2015distilling} and model compression \cite{buciluǎ2006model} technologies, enhancing MVAD performance on source artifacts such as \textit{graininess}, \textit{banding} and \textit{motion blur}, and improving the generalization of the model for an increasing number of different artifact types.

\noindent\textbf{Acknowledgement.} The authors appreciate the funding from Amazon Research Award, Fall 2022 CFP and the UKRI MyWorld Strength in Places Programme (SIPF00006/1), the University of Bristol. Any opinions, findings, conclusions or recommendations expressed in this material are those of the author(s) and do not reflect the views of Amazon.

\medskip

%%%%%%%%% REFERENCES
{\small
\bibliographystyle{ieee_fullname}
\bibliography{egbib}
}

\end{document}

%%%%%%%%% TITLE
\title{MVAD: A Multiple Visual Artifact Detector for Video Streaming\\ Supplementary Material}

\author{%
  Chen Feng$\dagger$, Duolikun Danier$\dagger$, Fan Zhang$\dagger$, Alex Mackin$\ddagger$, Andrew Collins$\ddagger$, David Bull$\dagger$ \\
  %\thanks{Use footnote for providing further information about author (webpage, alternative address)---\emph{not} for acknowledging funding agencies.} \\
  $\dagger$ Visual Information Lab, University of Bristol, BS1 5DD, United Kingdom \\
  $\ddagger$Amazon Prime Video, 1 Principal Place, Worship Street, London, EC2A 2FA, United Kingdom\\
  \texttt{\{chen.feng, duolikun.danier, fan.zhang, dave.bull\}@bristol.ac.uk}, \\ \texttt{\{acmackin, accllin\}@amazon.co.uk}
}

\maketitle
\thispagestyle{empty}
\pagestyle{empty}

\appendix

\section{ROC Plots}

\begin{figure*}[htbp]
\centering
    \includegraphics[width=0.325\linewidth]{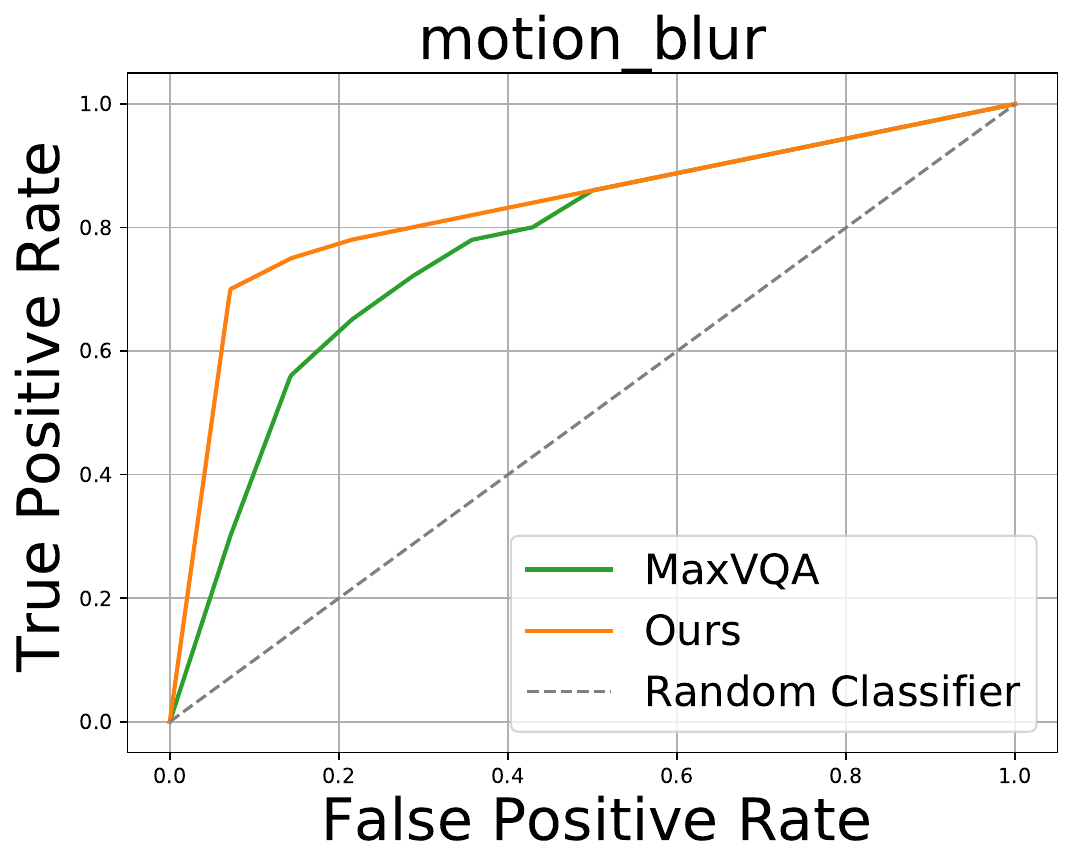}
    \includegraphics[width=0.325\linewidth]{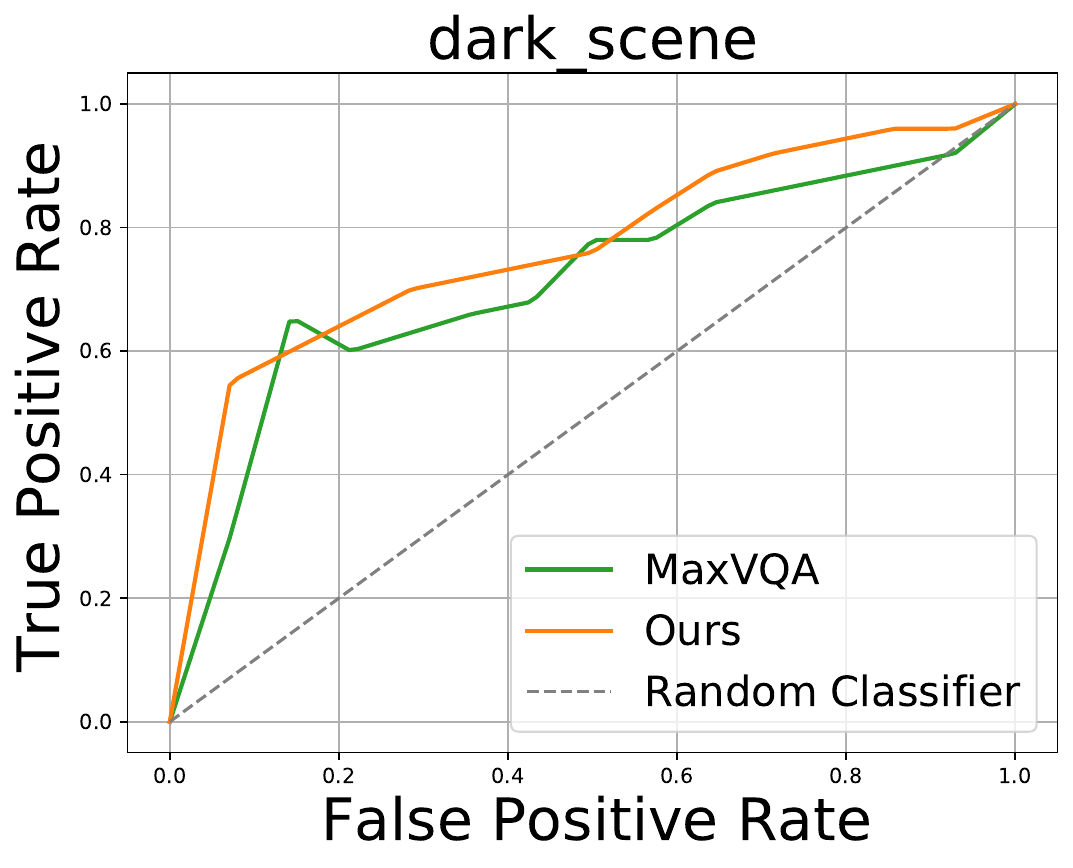}
    \includegraphics[width=0.325\linewidth]{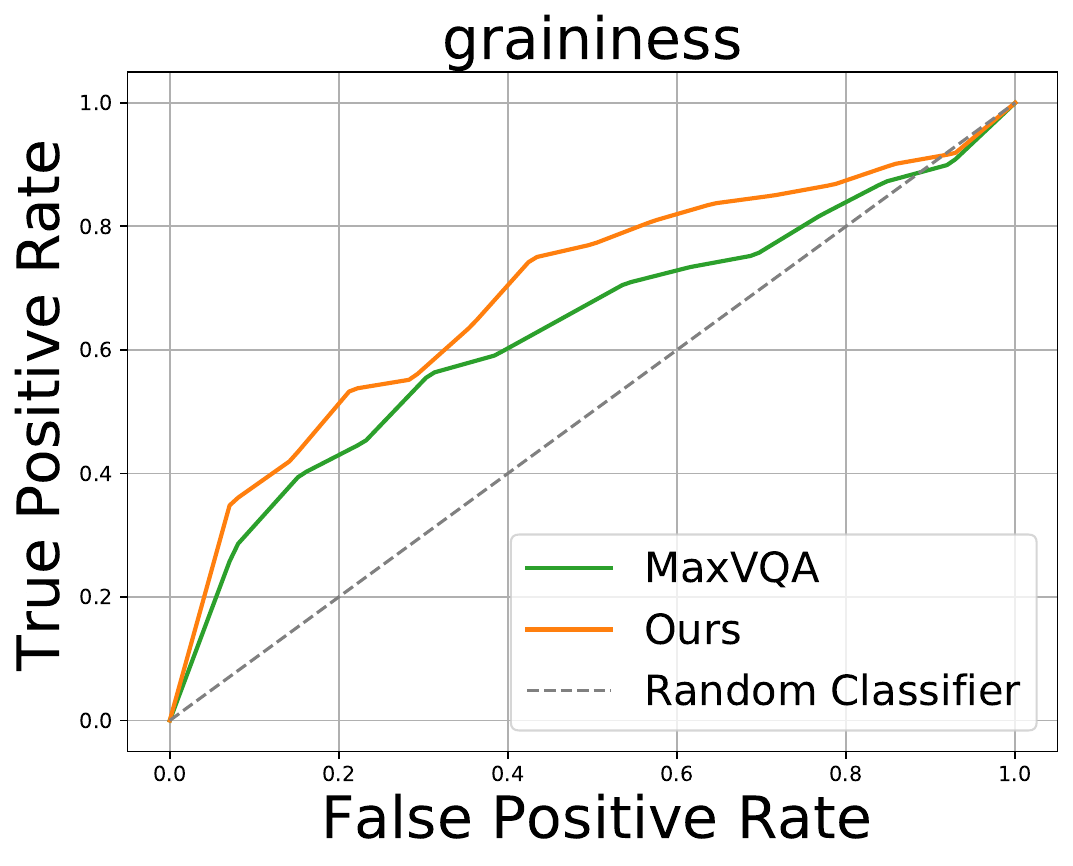}
    \includegraphics[width=0.325\linewidth]{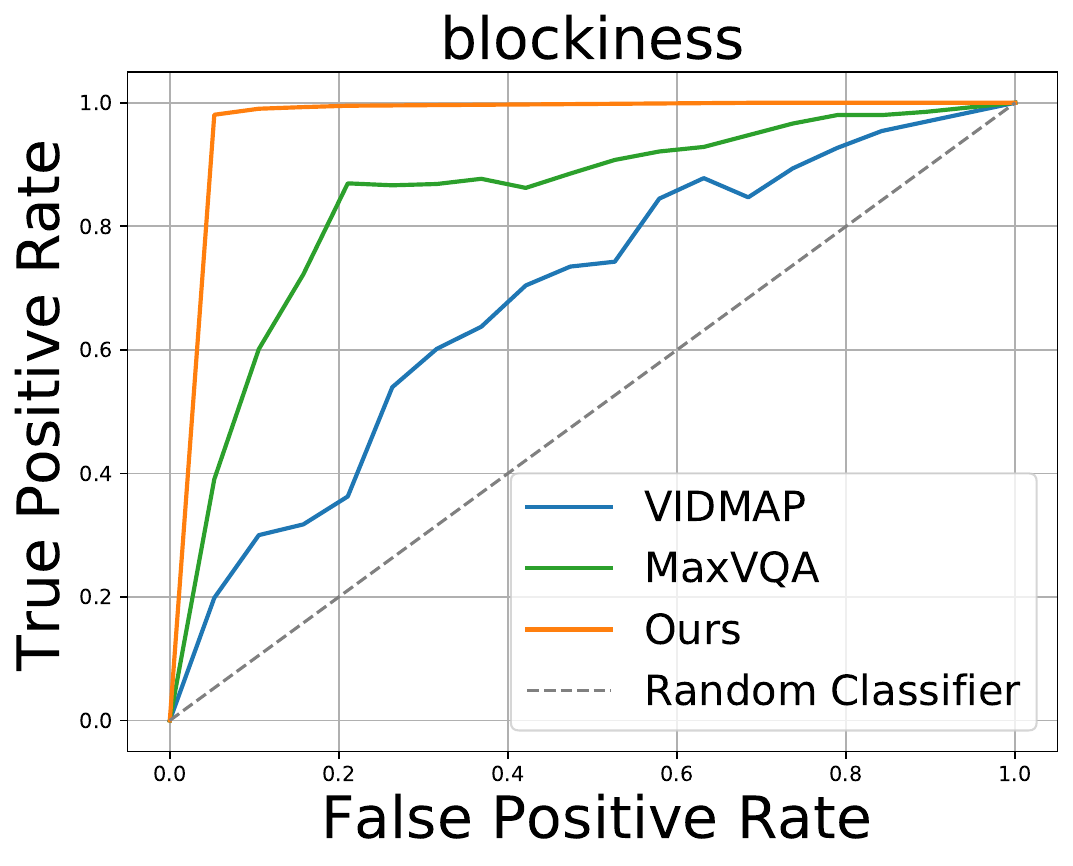}
    \includegraphics[width=0.325\linewidth]{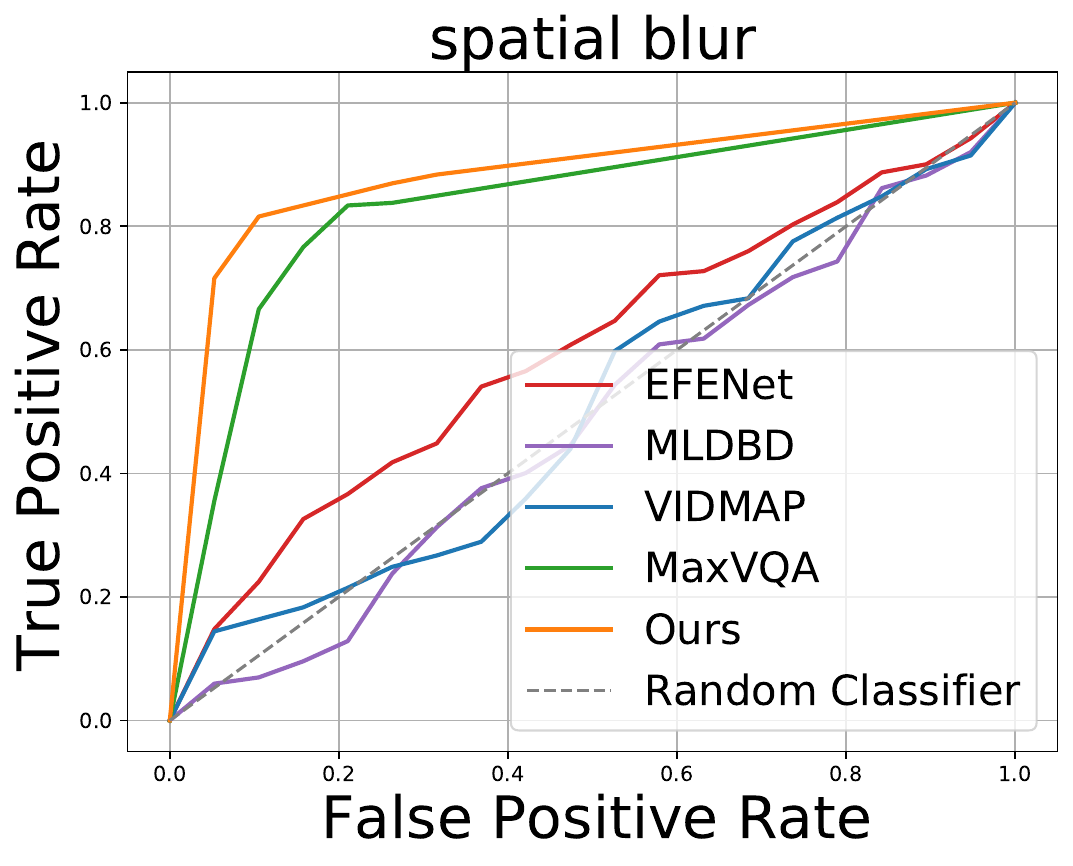}
    \includegraphics[width=0.325\linewidth]{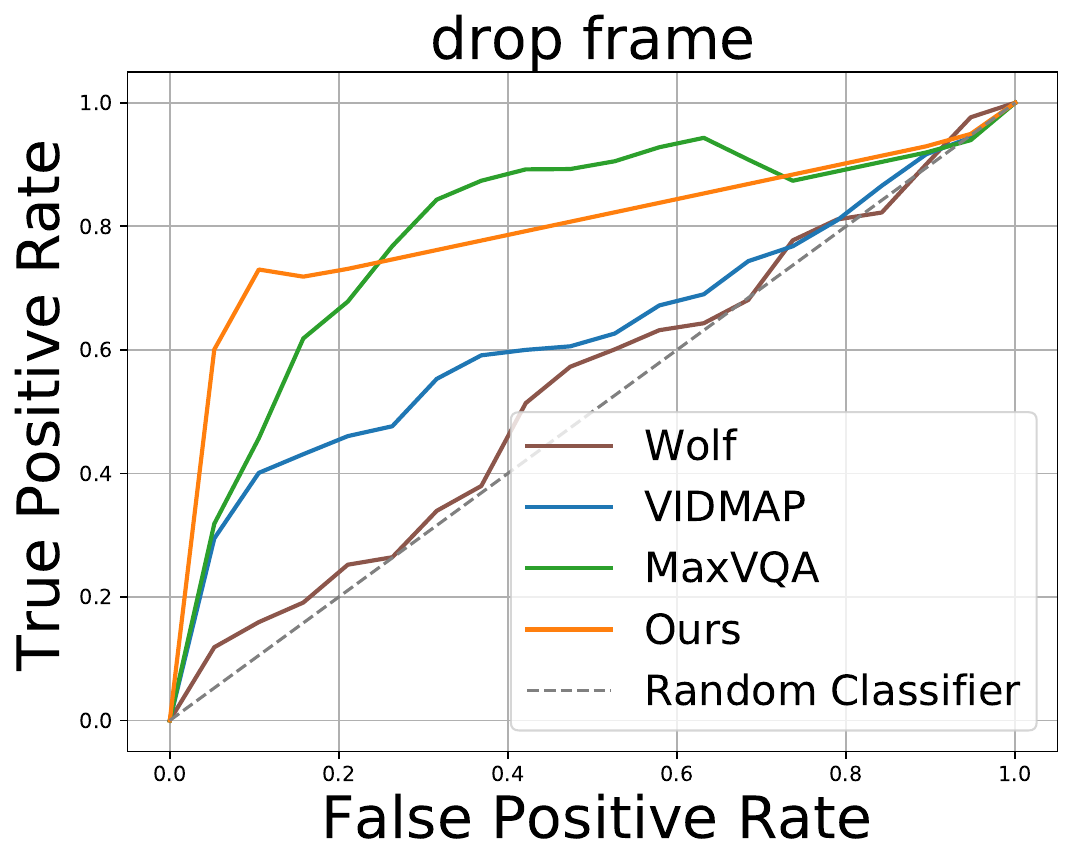}
    \caption{The ROC curves for different artifact categories in the Maxwell \cite{MaxVQA} database.}
    \label{fig:roc1}
\end{figure*}

\begin{figure*}[t!]
\centering
    \includegraphics[width=0.325\linewidth]{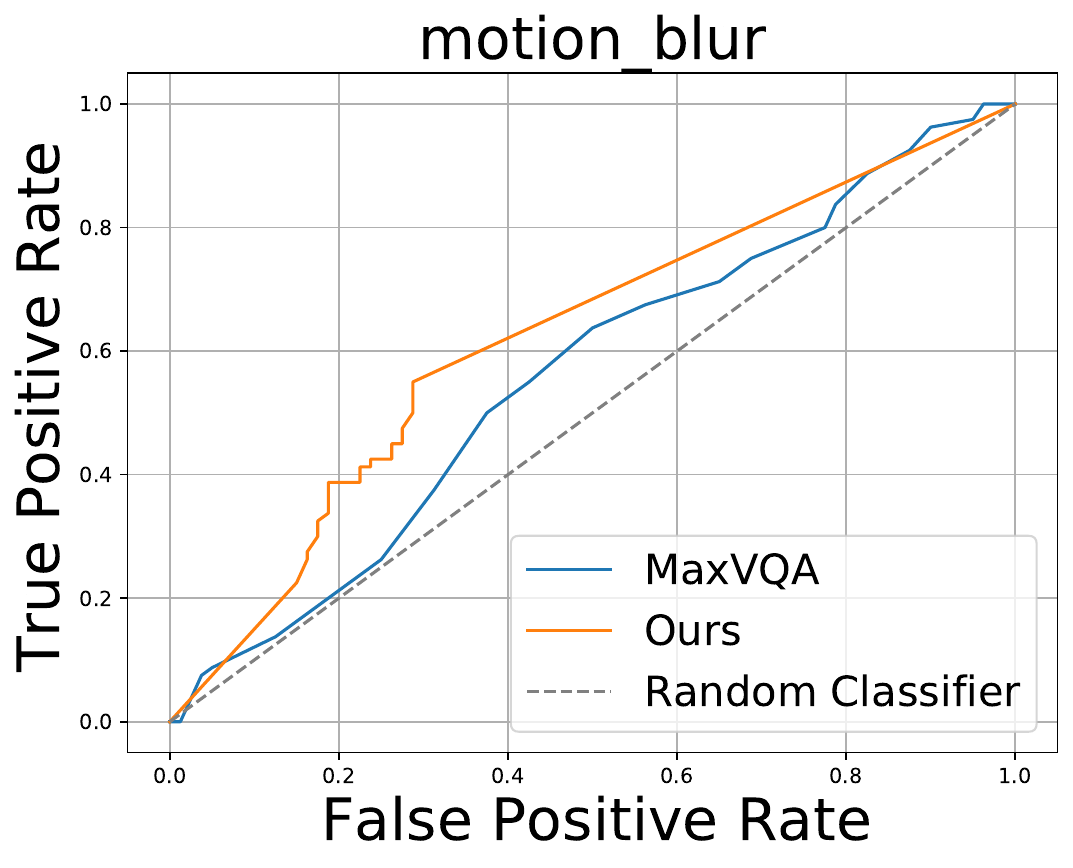}
    \includegraphics[width=0.325\linewidth]{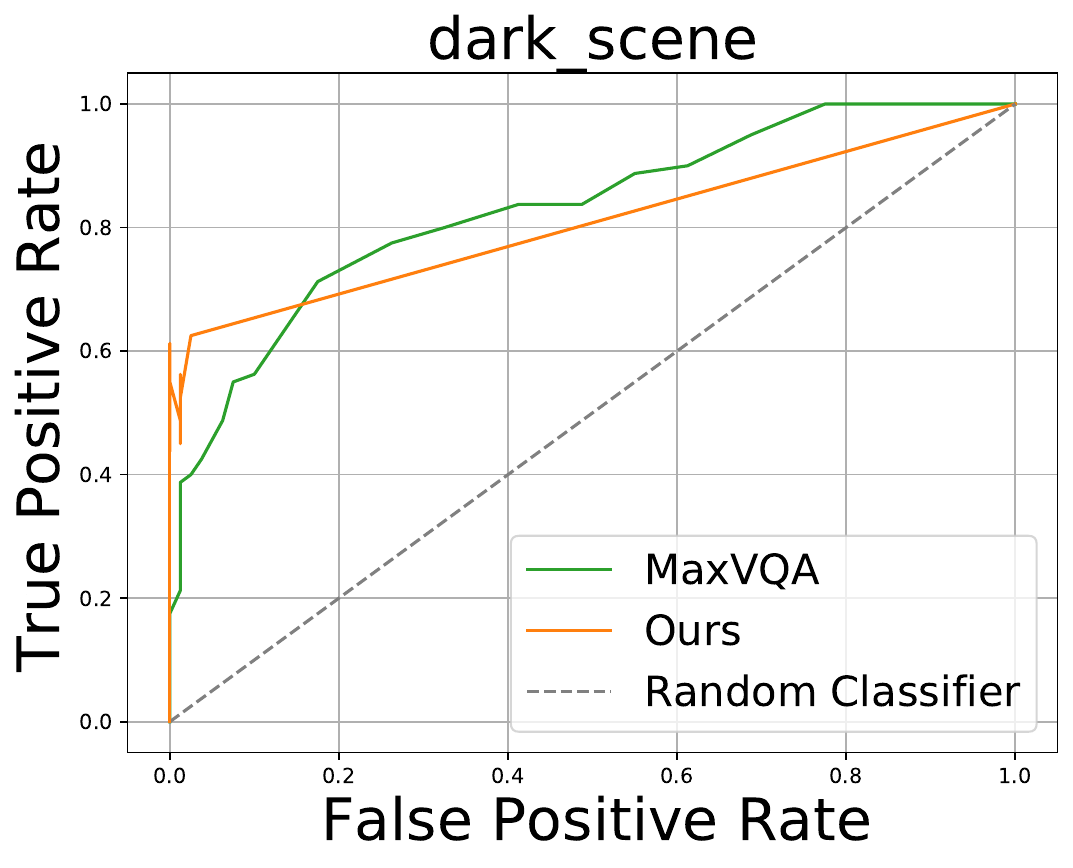}
    \includegraphics[width=0.325\linewidth]{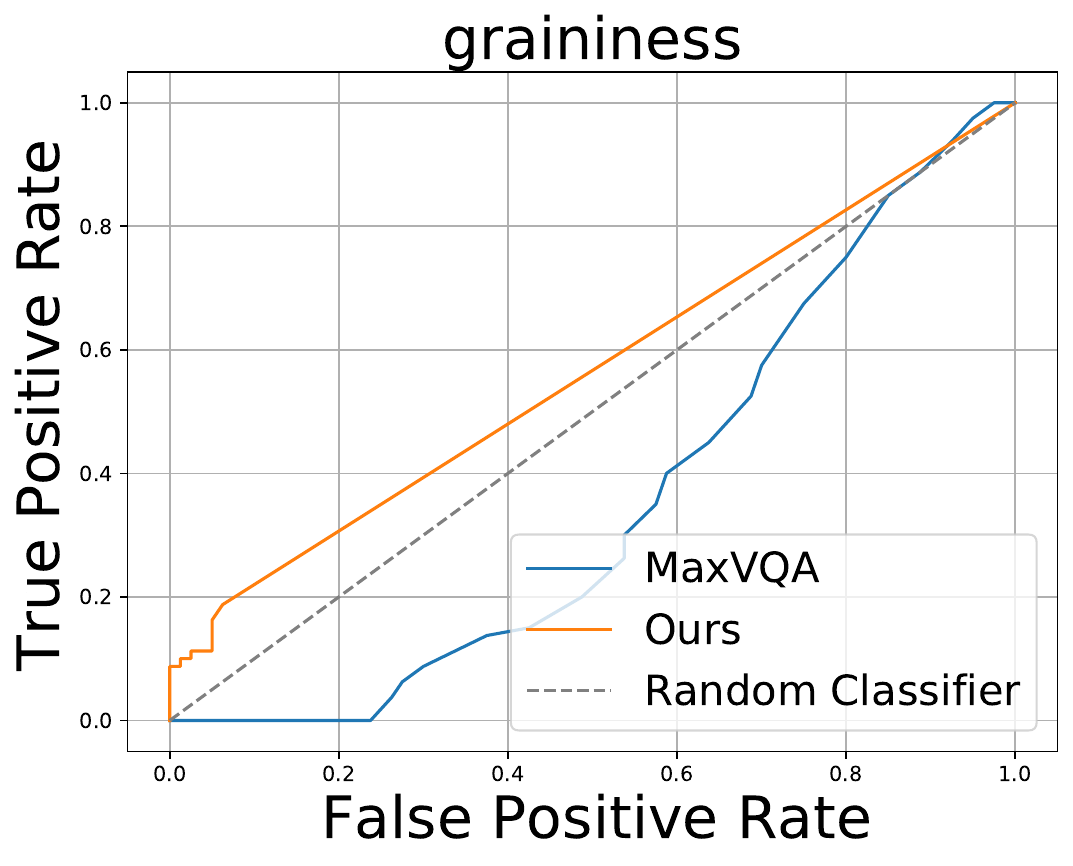}
    \includegraphics[width=0.325\linewidth]{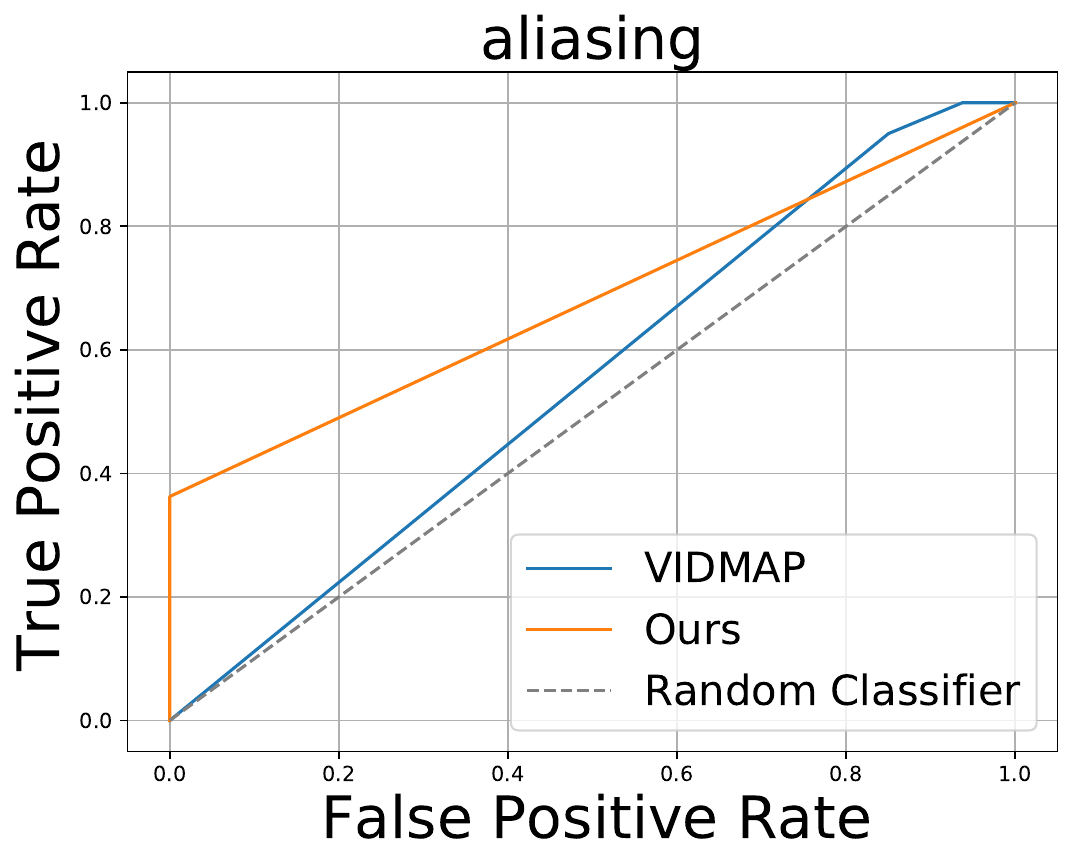}
    \includegraphics[width=0.325\linewidth]{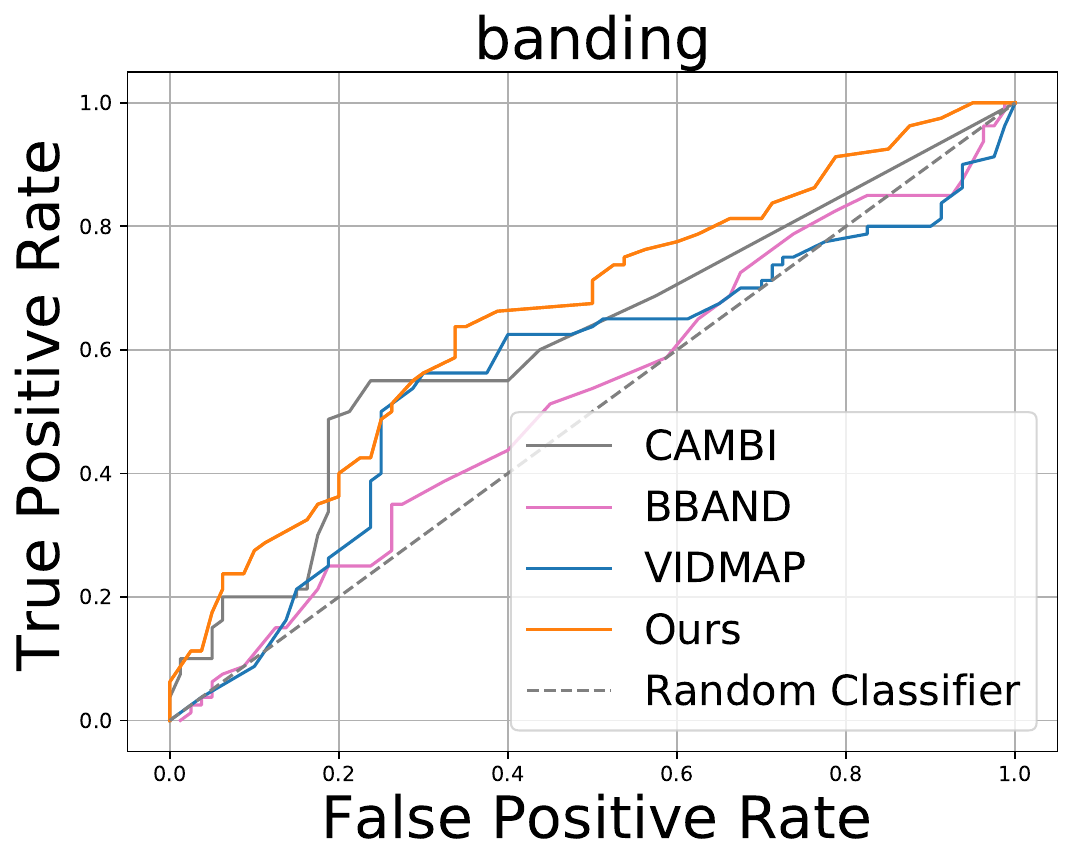}
    \includegraphics[width=0.325\linewidth]{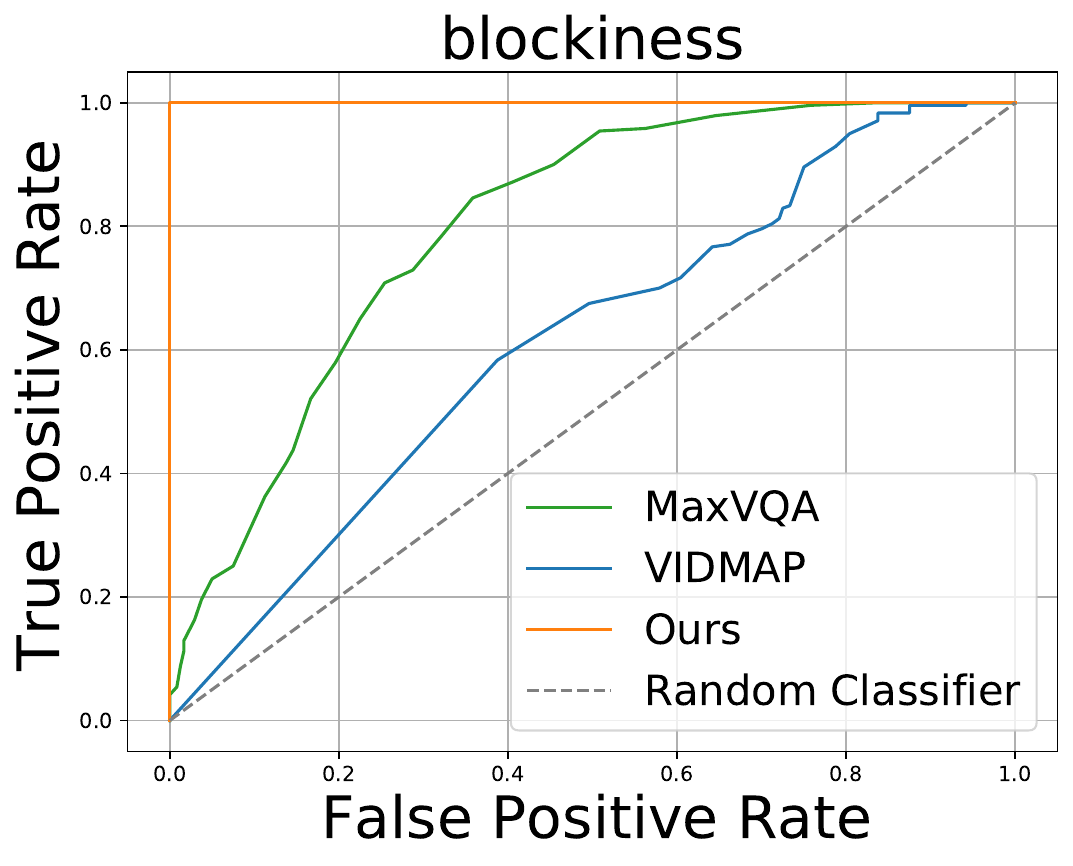}
    \includegraphics[width=0.325\linewidth]{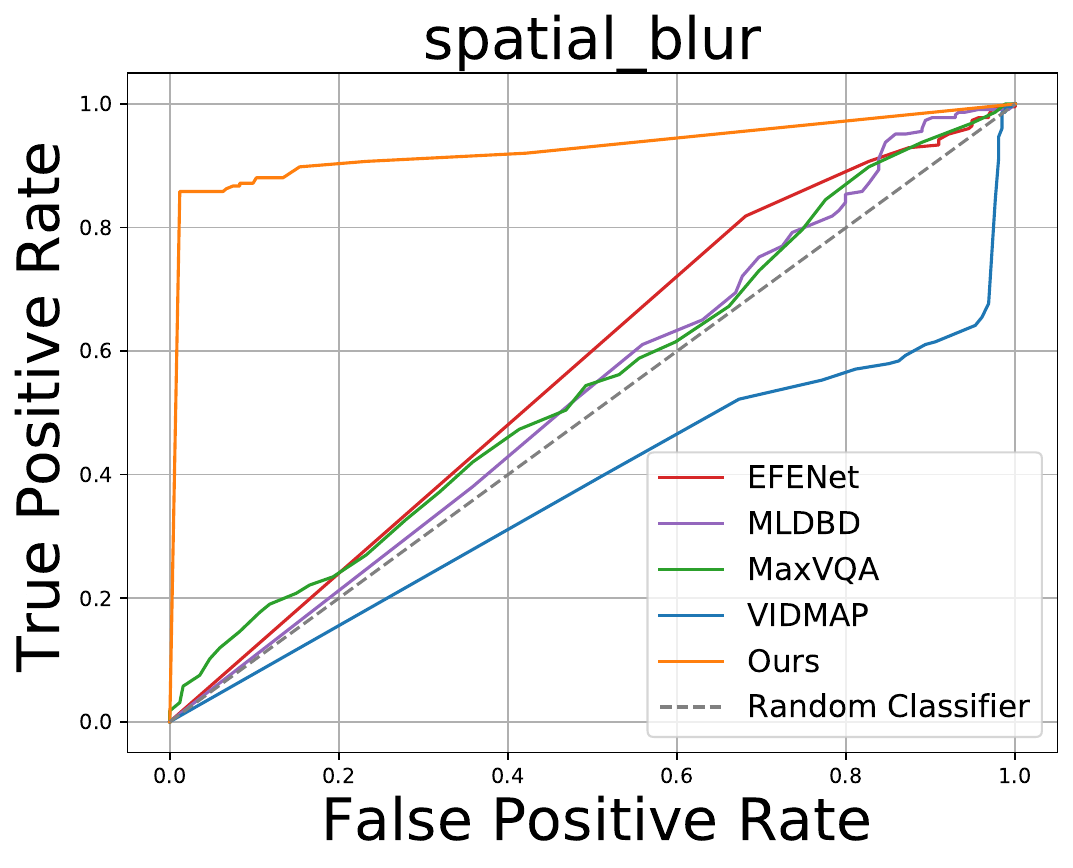}
    \includegraphics[width=0.325\linewidth]{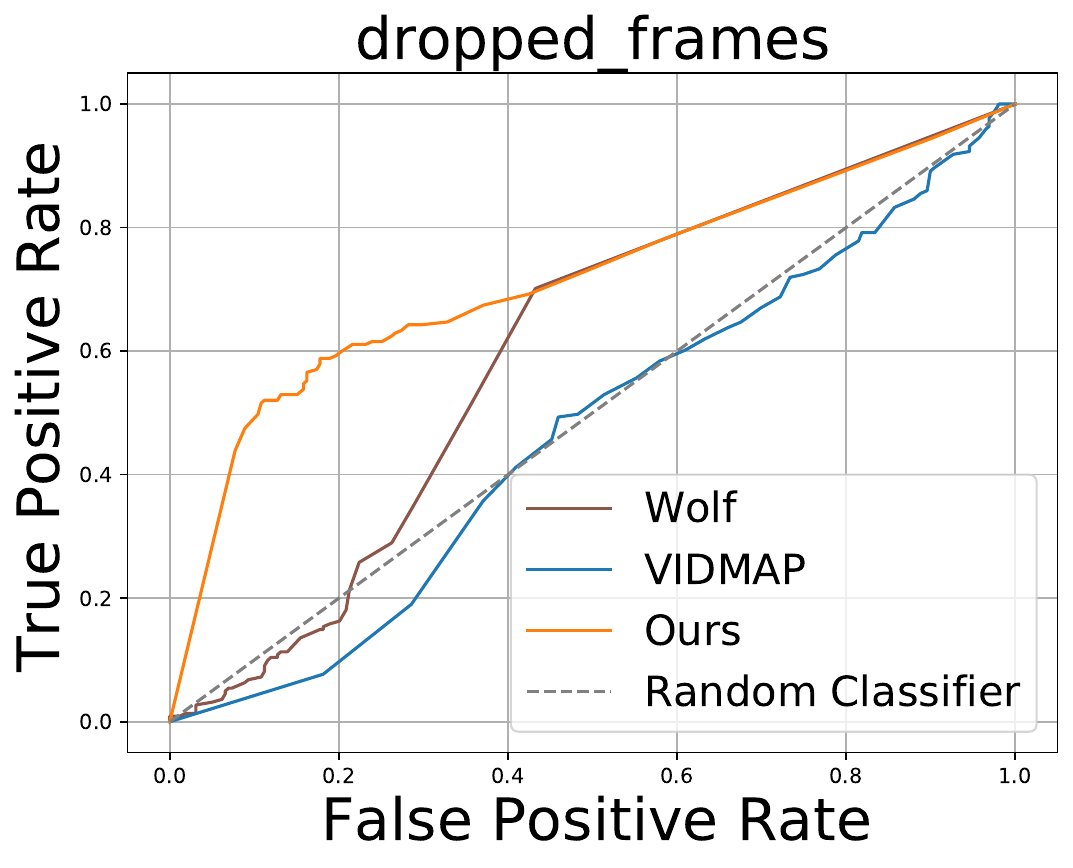}
    \includegraphics[width=0.325\linewidth]{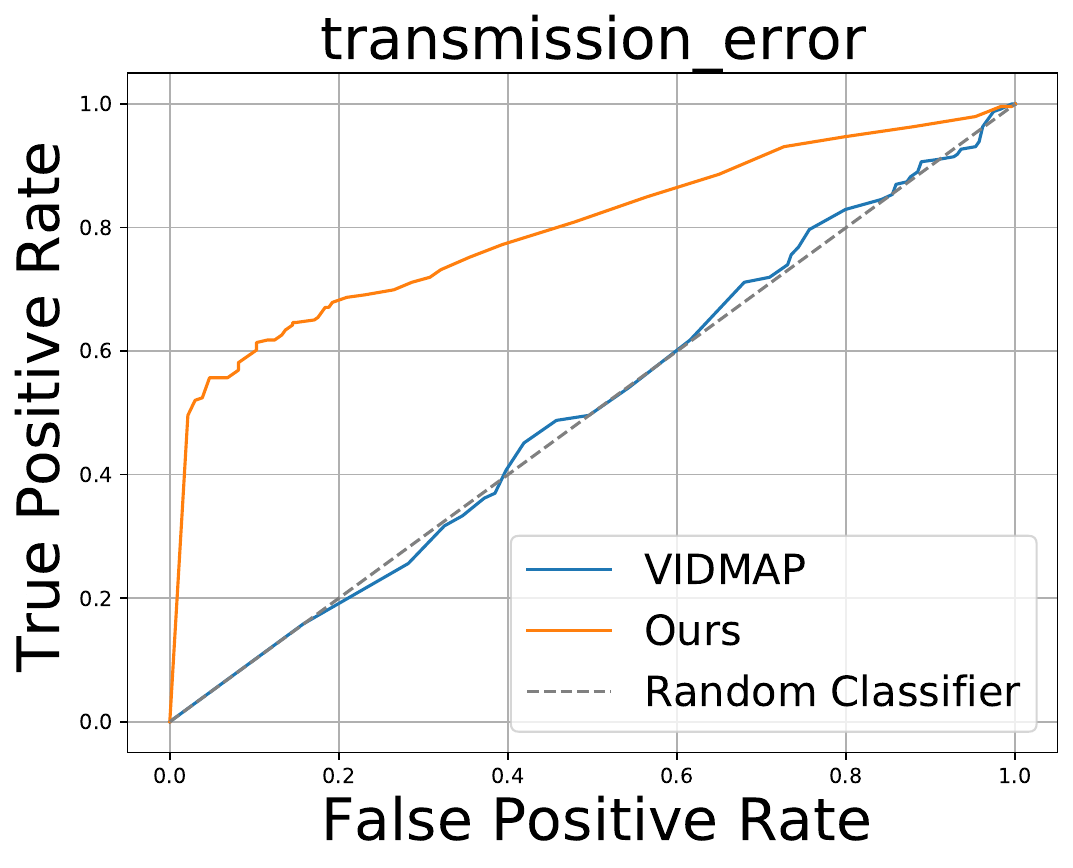}
    \includegraphics[width=0.325\linewidth]{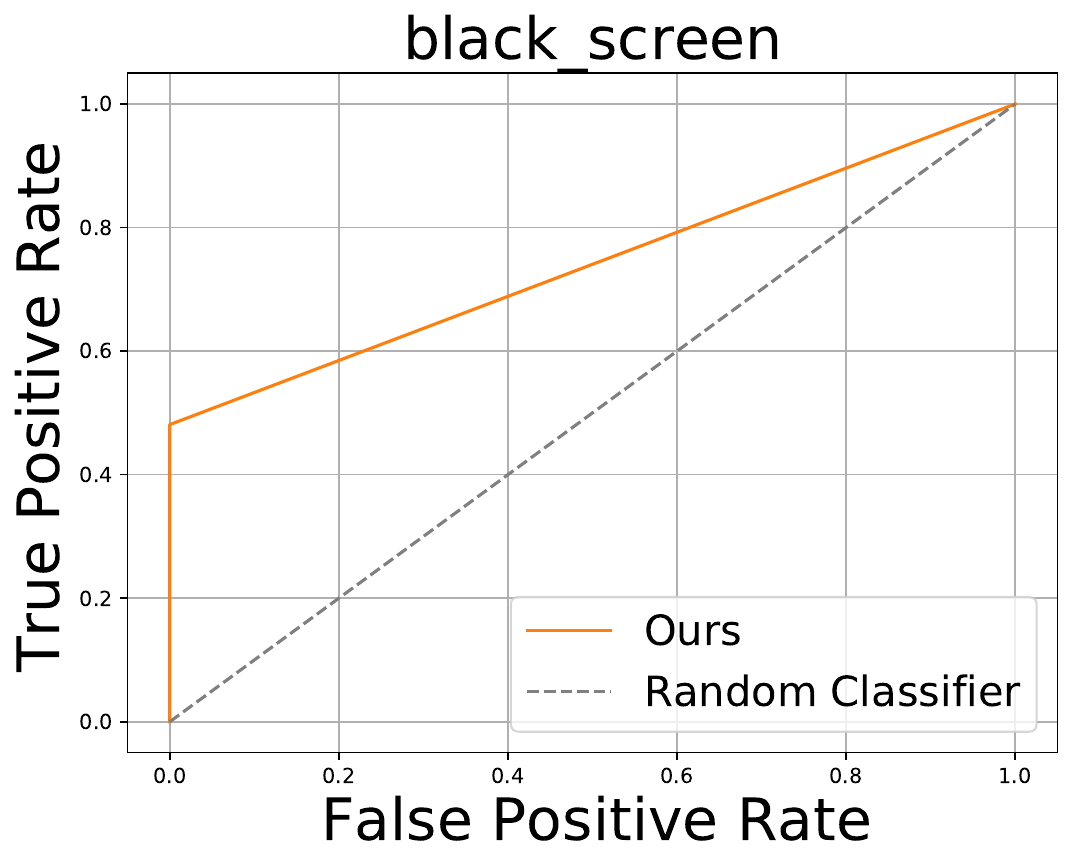}    
    \caption{The ROC curves for different artifact categories in the BVI-Artifact \cite{feng2023bvi} database.}
    \label{fig:roc2}
\end{figure*}

\autoref{fig:roc1} and \autoref{fig:roc2} show the ROC plots for eight artifact detection methods on each artifact tested based on the Maxwell and BVI-Artifact database. The ROC (Receiver Operating Characteristic) curves provide a visual representation of the performance of these methods, illustrating the trade-off between the true positive rate (TPR) and the false positive rate (FPR) across different threshold settings.

\section{Visualization of the Guided Mask in ADFE}

\begin{figure*}[htbp]
         \centering
        \centerline{\includegraphics[width=1\linewidth]{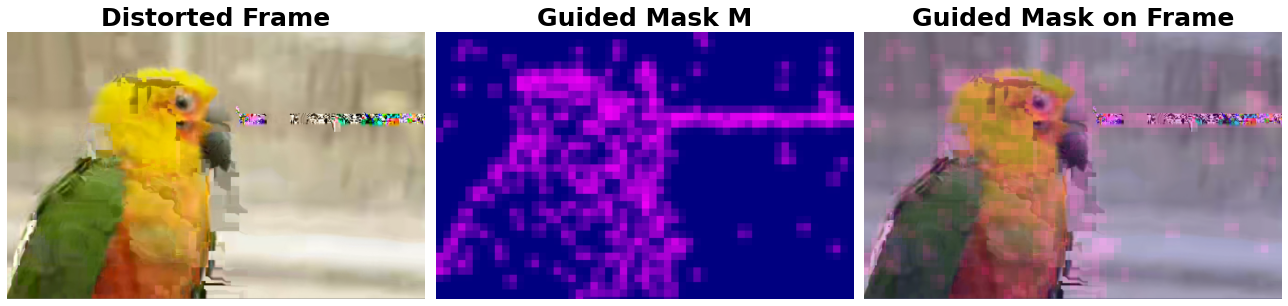}}
             \caption{Visualization of the guided mask $\mathbf{M}$ generated by the ADFE module.}
             \label{fig:map}
         \end{figure*}

\autoref{fig:map} showcases the influence of the guided mask $\mathbf{M}$ in the ADFE module, which does emphasize the regions with visual artifacts.

\section{Artifacts Synthesis Methods}

Table \ref{tab:artifacts_synthesis} provides a summary of the synthesis methods used to generate various video artifacts. In the Augmented Database, different parameters were employed to produce artifacts with four visibility levels.

\begin{table*}[t!]
    \centering
    \resizebox{0.85\linewidth}{!}{\begin{tabular}{r| l| r |c}
    \toprule
    Artifacts          & Synthesis Methods                                     & Levels                                           & Parameters \\
    \midrule
    \multirow{4}{*}{Aliasing}           & \multirow{4}{*}{Spatial re-sampling \cite{vidmap} }
        & Very noticeable& Sampling ratio=4\\
        && Noticeable & Sampling ratio=3\\
        && Subtle  & Sampling ratio=2\\
        && Very subtle   & Sampling ratio=1.5\\ \midrule
    \multirow{4}{*}{Banding}           & \multirow{4}{*}{Quantization \cite{vidmap}}  & Very noticeable& Quantization ratio=5\\
        && Noticeable & Quantization ratio=4\\
        && Subtle  & Quantization ratio=3\\
        && Very subtle   & Quantization ratio=2\\ \midrule
    \multirow{4}{*}{Dark Scene}         & \multirow{4}{*}{Brightness/contrast adjustment \cite{guo2016lime}} & Very noticeable& Decrease ratio=4\\
        && Noticeable & Decrease ratio=3\\
        && Subtle  & Decrease ratio=2\\
        && Very subtle   & Decrease ratio=1.5\\ \midrule
    \multirow{4}{*}{Motion Blur}        & \multirow{4}{*}{Convolution (motion blur) \cite{shen2020blurry}} & Very noticeable& Frame numbers=16\\
        && Noticeable & Frame numbers=12\\
        && Subtle  & Frame numbers=8\\
        && Very subtle   & Frame numbers=4\\ \midrule
    \multirow{4}{*}{Graininess}         & \multirow{4}{*}{Gaussian noise \cite{vidmap}}   
    & Very noticeable& standard deviation=50\\
        && Noticeable & standard deviation=25\\
        && Subtle  & standard deviation=10\\
        && Very subtle   & standard deviation=5\\ \midrule
    \multirow{4}{*}{Blockiness}         & \multirow{4}{*}{JPEG compression \cite{vidmap}}       & Very noticeable& Quantization parameter= 47\\
        && Noticeable & Quantization parameter= 42\\
        && Subtle  & Quantization parameter= 37\\
        && Very subtle   & Quantization parameter= 32\\ \midrule
    \multirow{4}{*}{Frame Drop}         & \multirow{4}{*}{Random frame dropping \cite{wolf2008no}}               & Very noticeable& Frame length = 16\\
        && Noticeable & Frame length = 12\\
        && Subtle  & Frame length = 8\\
        && Very subtle   & Frame length = 4\\ \midrule
    \multirow{4}{*}{Spatial Blur}       & \multirow{4}{*}{Convolution (Gaussian blur) \cite{vidmap}}                 & Very noticeable& Kernel size=9\\
        && Noticeable & Kernel size=7\\
        && Subtle  & Kernel size=5\\
        && Very subtle   & Kernel size=3\\ \midrule
    \multirow{4}{*}{Transmission Error} & \multirow{4}{*}{Packet loss and error concealment \cite{vidmap}}         & Very noticeable& bitstream filter = 4M\\
        && Noticeable & bitstream filter = 2M\\
        && Subtle  & bitstream filter = 1M\\
        && Very subtle   & bitstream filter = 0.5M\\ \midrule
    \multirow{4}{*}{Black Screen}       & \multirow{4}{*}{Random black frames replacement} & Very noticeable& Frame length = 16\\
        && Noticeable & Frame length = 12\\
        && Subtle  & Frame length = 8\\
        && Very subtle   & Frame length = 4\\ 
    \bottomrule
    \end{tabular}}
        \caption{Artifacts synthesis methods used for multiple artifacts generation. \label{tab:artifacts_synthesis}}
\end{table*}

\section{Broader Impacts}

The development and implementation of the MVAD framework for detecting multiple visual artifacts in streamed videos have the following impacts. 

On the positive side, the MVAD system supports automatic inspection of video content to identify visual artifacts, significantly reducing labor costs and increasing the efficiency of quality control processes in video streaming services. MVAD can also potential improve the quality of streamed videos by incorporating with video enhancement methods, leading to a better viewing experience for end users. 

The implementation of MVAD can also result in negative impacts. The training and deployment of sophisticated machine learning models such as MVAD require substantial computational resources, which can lead to increased energy consumption and negative environmental impact. Further model complexity reduction can alleviate this issue, and it remains our future work.

\section{License of Code and Data}

\autoref{tab:license} summarizes the license associated with the code and data used and generated in this work. 

\begin{table*}[t!]
\centering
\resizebox{\textwidth}{!}{\begin{tabular}{r ||  c| p{8cm} | c}
\toprule
Code/Data & Size & Dataset URL & License/Terms of Use \\
\midrule \midrule 
        \multicolumn{4}{c}{Training Datasets}\\
\midrule  BVI-DVC \cite{ma2021bvi} & 800 & \url{https://fan-aaron-zhang.github.io/BVI-DVC/} & Academic research.\\
\midrule  BVI-CC  \cite{katsenou2022bvi} & 90 &\url{https://fan-aaron-zhang.github.io/BVI-CC/} & Academic research.\\
\midrule  NFLX-public \cite{w:VMAF}&70&\url{https://github.com/Netflix/vmaf/blob/master/resource/doc/datasets.md}& Academic research.\\
\midrule LIVE-HFR \cite{madhusudana2021subjective}&88&\url{https://fan-aaron-zhang.github.io/BVI-HFR/}& Academic research.\\
\midrule Adobe240 \cite{su2017deep}&133&\url{https://www.cs.ubc.ca/labs/imager/tr/2017/DeepVideoDeblurring/}& MIT license.\\
\midrule\multicolumn{4}{c}{Test Datasets}\\
\midrule BVI-Artifact \cite{feng2023bvi} &480&\url{https://chenfeng-bristol.github.io/BVI-Artefact/}&Academic research.\\
\midrule Maxwell \cite{MaxVQA}&4,543&\url{https://github.com/VQAssessment/ExplainableVQA}& MIT license.\\
\midrule\multicolumn{4}{c}{Code of Benchmark Methods}\\
\midrule MaxVQA \cite{MaxVQA}&-&\url{https://github.com/VQAssessment/ExplainableVQA}& MIT license.\\
\midrule VIDMAP \cite{vidmap}&-&\url{https://github.com/utlive/VIDMAP}& Academic research.\\
\midrule CAMBI \cite{tandon2021cambi}&-&\url{https://github.com/Netflix/vmaf/blob/master/resource/doc/cambi.md}& BSD+Patent.\\
\midrule BBAND \cite{tu2020bband}&-&\url{https://github.com/google/bband-adaband}&Apache-2.0 license.\\
\midrule EFENet \cite{zhao2021defocus}&-&\url{https://github.com/wdzhao123/DENets}& No licence.\\
\midrule MLDBD \cite{zhao2023full} &-&\url{https://github.com/wdzhao123/MLDBD}& No licence.\\
\midrule Wolf \textit{el at.} \cite{wolf2008no}&-&\url{https://its.ntia.gov/umbraco/surface/download/publication?reportNumber=TR-09-456.pdf}& Academic research.\\\\
\midrule\multicolumn{4}{c}{Code and Data generated in this work}\\
\midrule  MVAD  & - & \url{https://chenfeng-bristol.github.io/MVAD/} & CC-BY-4.0 \\
\midrule  
Training data \cite{ma2021bvi}  & 50,800 & \url{https://chenfeng-bristol.github.io/MVAD/} & Academic research. \\
\bottomrule		
\end{tabular}}

\caption{License information for the Code and datasets used and generated in this work.}\label{tab:license}
\end{table*}

\clearpage
{\small
\bibliographystyle{ieee_fullname}
\bibliography{egbib}
}